\documentclass[manuscript]{aastex}
\usepackage{graphicx}
\usepackage{amsmath}

\usepackage{color}

\newcommand{\myemail}{georgakarakosl@hotmail.com}

\renewcommand{\vec}{\textbf}

\shorttitle{Analytic orbit propagation}
\shortauthors{Georgakarakos and Eggl}

\begin{document}

\title{ANALYTIC ORBIT PROPAGATION FOR TRANSITING CIRCUMBINARY PLANETS}

\author{Nikolaos Georgakarakos\altaffilmark{1}, Siegfried Eggl\altaffilmark{2}}
\email{\myemail}

\altaffiltext{1}{New York University Abu Dhabi, Saadiyat Island, PO Box 129188, Abu Dhabi, UAE}

\altaffiltext{2}{IMCCE, Observatoire de Paris, 77 Avenue Denfert-Rochereau,
F-75014, Paris, France}


\begin{abstract}
The herein presented analytical framework fully describes the motion of coplanar systems consisting of a stellar binary and a planet orbiting both stars
on orbital as well as secular timescales.  Perturbations of the Runge-Lenz vectors are used to derive short period evolution of the system, while octupole 
secular theory is applied to describe its long term behaviour. A post Newtonian correction on the stellar orbit is included.  The planetary orbit is initially circular 
and the theory developed here assumes that the planetary eccentricity remains relatively small ($e_2<0.2$). 
Our model is tested against results from numerical integrations of the full equations of motion and is then applied 
to investigate the dynamical history of some of the circumbinary planetary systems discovered by NASA's Kepler satellite.
Our results suggest that the formation history of the systems Kepler-34 and Kepler-413 has most likely been different from the one of 
Kepler-16, Kepler-35, Kepler-38 and Kepler-64, since the observed planetary eccentricities for those systems are not compatible with the assumption of initially circular orbits. 

\end{abstract}


\section{INTRODUCTION}
Many problems in dynamical astronomy and celestial mechanics can be reduced to describing gravitational interactions in hierarchical triple systems.
Examples range from asteroid \citep[e.g.][]{Liu-et-al-2012} and planetary systems \citep[e.g.][]{Teyssandier-et-al-2013} to 
multi star configurations \citep[e.g][]{Borkovits-et-al-2007} or even systems with compact objects such as black holes \citep[e.g.][]{Blaes-et-al-2002}.
Hierarchical triples consist of a binary system and a third body on a
wider orbit. The dynamical behavior of such a configuration can be pictured as the motion
of two binaries: the two inner bodies (inner binary) move on a perturbed Keplerian orbit, whereas the third body with the centre of mass of the inner binary form
the so-called outer binary. Both orbits exhibit coupled dynamical evolution.

The dynamical behaviour of hierarchical systems, especially the long term (secular) behaviour, has been a subject of intense study in the past. 
 \citet{Brouwer-1959}, \citet{Kaula-1962}, \citet{Kozai-1962}, \citet{1962P&SS....9..719L}, \citet{Harrington-1968} and \citet{Heppenheimer-1978} are excellent
examples of that.  Research on the subject has continued throughout the years with \citet{Krymolowski-et-al-1999}, 
\citet{Ford-et-al-2000}, \citet{Lee-et-al-2003}, \citet{Farago-et-al-2010}, \citet{Naoz-et-al-2013a, Naoz-et-al-2013b}, \citet{2013ApJ...763..107L}, \citet{2014ApJ...791...86L} and \citet{2015MNRAS.447..747L} being a small sample of recent developments in the subject. 

If the system to be modeled is coplanar, then the dynamical evolution of the hierarchical triple is dominated by the 
time dependent changes in the eccentricity vectors (Laplace-Runge-Lenz vectors) of the inner and outer orbit, as the semi-major axes do not show any secular evolution \citep[e.g.][]{Marchal-1990}. Hence, a solution of the differential equations
governing the behaviour of the eccentricity vectors allows us to accurately describe the motion of the whole system via closed analytic formulae.

In a series of papers, we focused on the evolution of the inner eccentricity of a hierarchical triple system.  
In \citet{Georgakarakos-2002}, we developed a method for measuring the eccentricity injected to the inner orbit 
due to the perturbations of the third body. This effect has been known for years \citep[e.g.][]{Mazeh-1979}, but
\citet{Georgakarakos-2002} quantified it analytically on both short term and secular timescales for the first time. As those calculations were performed for 
coplanar and initially circular orbits, they were later on extended to cases where the perturber was
on an eccentric orbit \citep{Georgakarakos-2003} as well as for three dimensional cases \citep{Georgakarakos-2004}. The initial
motivation for that work was closely related to modeling tidal friction in stellar triple systems \citep[e.g. see][]{Kiseleva-et-al-1998}. Therefore, the systems investigated 
had comparable masses.  Nonetheless, the derived formulae were also tested numerically for planetary mass ratios
and remained valid \citep{Georgakarakos-2006}.  Finally, an improved estimates in the case of coplanar and initially circular orbits were
given in \citet{Georgakarakos-2009}. The analytical formulae for the eccentricity evolution in hierarchical triples have proven to be quite versatile. \citet{Chavez-2012} used them
to investigate the long term modulation in the lightcurve of the cataclysmic variable FS Aurigae, which was
suspected to happen due to the gravitational perturbation of a third body on a much wider orbit. More recently, the limits of various types of habitable zones for a planet in a stellar binary with different spectral type components were investigated analytically as well as numerically \citep{Eggl-et-al-2012, Eggl-et-al-2013}. 

Boosted by the discovery of several circumbinary planets \citep[e.g.][]{2011Sci...333.1602D, Welsh-et-al-2012, 2012ApJ...758...87O, 2013ApJ...768..127S, 2014ApJ...784...14K}, the interest in analytical models that describe cicumbinary planetary motion has recently grown.  In order to create more accurate analytical tools for detection algorithms and planetary formation theories we have to improve first and foremost our description of the eccentricity evolution of planets on circumbinary orbits (also known as P-type orbits).  There has been
some progress in \citet{Georgakarakos-2009} regarding the latter matter, but only hierarchical triple systems on initially circular orbits were treated therein, which may be inadequate in certain cases.  The Kepler-34 system for example has a stellar binary with an eccentricity of ${e \approx 0.52}$ \citep{Welsh-et-al-2012}.

In this paper, we derive analytical expressions for the eccentricity of a relatively light body orbiting a heavier binary, e.g. a 
planet moving around an eccentric double star. This will allow us to model the dynamical evolution of a circumbinary planet analytically. 
Since most of the circumbinary planets that have been found via transit searches share more or less the same orbital plane, we will
focus on coplanar configurations. Furthermore, we include a post-Newtonian correction into our model, since, in certain cases, General Relativity (GR) may affect the orbit of the inner binary, which in turn influences the evolution of the planetary eccentricity.  Our model assumes that the planetary eccentricity is initially circular and it remains valid for low values of the planet's eccentricity ($e_2<0.2$).

The remainder of this article is structured as follows: in section \ref{sec:ecc}, we present the analytical derivation for the eccentricity of the
planetary orbit, as well as the expressions for its maximum and averaged value.  In section \ref{sec:test}, the analytical estimates are compared to results obtained
from the numerical integration of the full equations of motion.  
Next, we discuss the effect of a post-Newtonian correction in our problem (section \ref{sec:PN}) and we
present the formulae for the analytic propagation of the hierarchical triple system (section \ref{sec:ana}). 
In section \ref{sec:kepler}, we apply our analytical estimates on six circumbinary planetary systems discovered during the Kepler mission.  
Finally, we discuss and summarize our results in section \ref{sec:summary}.  A brief description of the variables used in this work can be found in the notation section after the appendix.

\section{THEORY}
\label{sec:ecc}
We are interested in finding a complete description of the eccentricity vectors of the inner and outer orbit, since those will determine 
the dynamical evolution of a coplanar hierarchical triple system.  As it was stated earlier, we will
assume a star-star planet configuration, where all bodies are treated as point-masses.
Since the outer body is of planetary size, we can neglect its influence on the amplitude of orbital eccentricity of the inner pair (this will be justified in section \ref{sec:test}).  Only changes in the direction of the eccentricity vector of the inner orbit will be taken into consideration.

The general way to derive accurate eccentricity estimates that contain long and short periodic contributions has been laid out in \citet{Georgakarakos-2003}.
First, the long term evolution of the eccentricity vector (Laplace-Runge-Lenz vector) is studied by deriving and solving equations of motion
that have been averaged over all fast angles.
The averaging process via a canonical transformation eliminates all short periodic terms in the eccentricity evolution, so that they do not appear in the final solution.  If we want to describe the system's behavior on orbital timescales as well, we can reintroduce short periodic terms by deriving the non-averaged equations for the eccentricity vector.
We then eliminate all the components that are already contained in the long term (secular) solution and solve the reduced differential equations.
The final solution is a superposition of the short and long periodic solutions of the eccentricity vector.

That kind of approach has been chosen in this work. The secular solutions of the outer eccentricity vector have been constructed from the equations of motion derived from 
the corresponding canonically averaged 
Hamiltonians. The short periodic terms, however, had to be themselves split into medium (with periods of the order of the planetary orbital period) and short term (with periods of the order of the stellar orbital period) components in order to allow for an analytic 
solution.
Rejoining the different contributions of the long, medium and short term solutions leads to the final result as detailed in the 
following sections.

\subsection{Calculation of the long term evolution}
\label{secular}

In order to derive the long-term modulation of the system, we shall use the Hamiltonian
formulation \citep{Marchal-1990}.  For this purpose, we make use of the Delaunay variables, which is a set of canonical variables. For 
a coplanar three body system they are defined as follows:
\begin{eqnarray}
\begin{split}
L_1=&mn_1a^{2}_{1}, \hspace{1cm} &l_1,\\
L_2=&\mathcal Mn_2a^{2}_{2}, \hspace{1cm} &l_2,\\
G_1=&L_{1}\sqrt{1-e^2_1}, \hspace{1cm} &g_1=\varpi_1,\\
G_2=&L_{2}\sqrt{1-e^2_2}, \hspace{1cm} &g_2=\varpi_2,\label{eq:delaunay}
\end{split}
\end{eqnarray}
where $a_i$, $e_i$, $\varpi_i$ and ${l_i}$ are the semi major axes, the eccentricities, the longitudes of pericenter and mean anomalies of the inner ($i=1$) and the outer ($i=2$) orbit respectively.
Furthermore,  
\begin{displaymath}
m=\frac{m_0m_1}{m_0+m_1}\hspace{0.3cm} \mbox{and} \hspace{0.3cm}\mathcal M =\frac{m_2(m_0+m_1)}{M}, 
\end{displaymath}
are the so-called reduced masses, with $m_0, m_1$ and $m_2$, being the masses of the stellar binary and planet respectively. The total mass of the system is $M=\sum_{i=0}^2m_i$.
Using those variables, the Hamiltonian for a hierarchical triple system reads 
\begin{equation}
\label{ham}
H=H_0+H_1+H_{p},
\end{equation}
where
\begin{equation}
H_0=-\frac{\mathcal{G}^2m^3_0m^3_1}{2(m_0+m_1)L^{2}_{1}}
\end{equation}
is the Keplerian energy of the inner orbit,
\begin{equation}
H_1=-\frac{\mathcal{G}^2(m_0+m_1)^3m^3_2}{2ML^{2}_{2}}
\end{equation}
is the Keplerian energy of the outer orbit, and
\begin{equation}
H_{p}=\mathcal{G}m_2(\frac{m_0+m_1}{R}-\frac{m_0}{r_{02}}-\frac{m_1}{r_{12}}), 
\end{equation}
is the perturbing Hamiltonian, with ${r_{02}}$ and ${r_{12}}$ being the distances between ${m_{0}}$ and ${m_{2}}$
and  ${m_{1}}$ and ${m_{2}}$ respectively.  
Furthermore, $\textit{\textbf{R}}$ is the vector which connects the centre of mass of the inner binary with the third body, while $\textit{\textbf{r}}$ will be denoting the relative position vector of the inner orbit (see Figure \ref{fig:jacobi}).  Finally, $\mathcal{G}$ is the gravitational constant.

Since we deal with a hierarchical system and ${r/R}$ is small, ${H_{p}}$
can be expressed in terms of Legendre polynomials, giving:
\begin{equation}
H_p=-\frac{\mathcal{G}m_0m_1m_2}{R}\sum^{\infty}_{j=2}M_j(\frac{r}{R})^j\mathcal{P}_j(\cos\theta),
\end{equation}
where ${M_j=\frac{m^{j-1}_{0}-(-m_1)^{j-1}}{(m_0+m_1)^j}}$, ${\mathcal{P}_j}$ are the Legendre polynomials and $\theta$, as seen in Figure \ref{fig:jacobi}, is the angle between the 
$\textit{\textbf{r}}$ and $\textit{\textbf{R}}$ vectors.

The long term behaviour of the system can be obtained when the short period effects in the Hamiltonian are removed
by using the Von Zeipel method \citep[e.g.][]{Marchal-1990}.  The method employes a canonical transformation and details about the derivation
can be found  elsewhere \citep{Brouwer-1959, Marchal-1990, Krymolowski-et-al-1999}.  The advantage of the Von Zeipel transformation 
compared to other averaging methods, such as for example the 'scissors method', is that it produces a Hamiltonian which is second order in terms of the masses (although that term will not be used here as its effect is negligible due to the fact that the third body has a planetary mass) and which is complete in terms of the eccentricities.  Hence, the Hamiltonian
for coplanar orbits averaged over the inner and outer mean anomalies is
\begin{equation}
<H>=<H_0>+<H_1>+<H_{p2}>+<H_{p3}>+\mathcal{O}(\frac{r^4}{R^5}), \label{hamilto}
\end{equation}
with
\begin{eqnarray}
\begin{split}
<H_0> &=-\frac{\mathcal{G}^2m^3_0m^3_1}{2(m_0+m_1)L^{2}_{1l}}\\
<H_1>&=-\frac{\mathcal{G}^2(m_0+m_1)^3m^3_2}{2ML^{2}_{2l}}\\
<H_{p2}> & =  -\frac{1}{8}\frac{\mathcal{G}^2(m_{0}+m_{1})^7m^7_{2}}{m^3_{0}m^3_{1}M^3}\frac{L^4_{1l}}{L^3_{2l}G^3_{2l}}(5-3\frac{G^2_{1l}}{L^2_{1l}})\\
<H_{p3}> & =  \frac{15}{64}\frac{\mathcal{G}^2(m_0+m_1)^9m^9_2(m_0-m_1)}{m^5_0m^5_1M^4}\frac{L^6_{1l}}{L^3_{2l}G^5_{2l}}
\sqrt{1-\frac{G^2_{1l}}{L^2_{1l}}}\sqrt{1-\frac{G^2_{2l}}{L^2_{2l}}}\times\\
&  \times\big(7-3\frac{G^2_{1l}}{L^2_{1l}}\big)\cos{(g_{1l}-g_{2l})}  \label{eq:hav}
\end{split}
\end{eqnarray}
where the index ${l}$ stands for the long term (secular) evolution and $H_{p2}$ and and $H_{p3}$ come from the $\mathcal{P}_2$ and $\mathcal{P}_3$ Legendre polynomials respectively.

Using Delaunay variables, Hamilton's equations assume the following form for a coplanar system:
\begin{eqnarray}
\begin{split}
\frac{dL_{jl}}{dt}=-\frac{\partial <H>}{\partial l_{jl}},& \hspace{0.5cm} \frac{dl_{jl}}{dt}=\frac{\partial <H>}{\partial L_{jl}}\\
\frac{dG_{jl}}{dt}=-\frac{\partial <H>}{\partial g_{jl}},& \hspace{0.5cm} \frac{dg_{jl}}{dt}=\frac{\partial <H>}{\partial G_{jl}}, \hspace{0.5cm}j=1,2.
\end{split}
\end{eqnarray}
Since our Hamiltonian is independent of ${l_{jl}}$, all ${L_{jl}}$ are constant. This is a well known result for the averaged probelm \citep[e.g.][]{Harrington-1968}.  Then, keeping in mind that the eccentricity of the stellar orbit remains unchanged due to
the fact that the outer body has a planetary mass, the relevant equations of motion are:  
\begin{eqnarray}
\frac{dG_{2l}}{dt} & = & -\frac{\partial <H_{p3}>}{\partial g_{2l}}=-\frac{15}{64}\frac{\mathcal{G}^2(m_0+m_1)^9m^9_2(m_0-m_1)}{m^5_0m^5_1M^4}\frac{L^6_{1l}}{L^3_{2l}G^5_{2l}}
\sqrt{1-\frac{G^2_{1l}}{L^2_{1l}}}\times\nonumber\\
& & \times\sqrt{1-\frac{G^2_{2l}}{L^2_{2l}}}\big(7-3\frac{G^2_{1l}}{L^2_{1l}}\big)\sin{(g_{1l}-g_{2l})}\\
\frac{dg_{2l}}{dt} & = & \frac{\partial <H_{p2}+H_{p3}>}{\partial G_{2l}}=\frac{3}{8}\frac{\mathcal{G}^2(m_{0}+m_{1})^7m^7_{2}}{m^3_{0}m^3_{1}M^3}\frac{L^4_{1l}}{L^3_{2l}G^4_{2l}}(5-3\frac{G^2_{1l}}{L^2_{1l}})-\nonumber\\
& & -\frac{15}{64}\frac{\mathcal{G}^2(m_0+m_1)^9m^9_2(m_0-m_1)}{m^5_0m^5_1M^4}\frac{L^6_{1l}}{L^3_{2l}}
\sqrt{1-\frac{G^2_{1l}}{L^2_{1l}}}\big(7-3\frac{G^2_{1l}}{L^2_{1l}}\big)\times\nonumber\\
& & \times\Bigg(\frac{5\sqrt{1-\frac{G^2_{2l}}{L^2_{2l}}}}{G^6_{2l}}+\frac{1}{L^2_{2l}G^4_{2l}
\sqrt{1-\frac{G^2_{2l}}{L^2_{2l}}}}\Bigg)\cos{(g_{1l}-g_{2l})}\\ 
\frac{dg_{1l}}{dt} & = & \frac{\partial <H_{p2}+H_{p3}>}{\partial G_{1l}}=\frac{3}{4}\frac{\mathcal{G}^2(m_{0}+m_{1})^7m^7_{2}}{m^3_{0}m^3_{1}M^3}\frac{L^2_{1l}G_{1l}}{L^3_{2l}G^3_{2l}}+\nonumber\\
& & -\frac{15}{64}\frac{\mathcal{G}^2(m_0+m_1)^9m^9_2(m_0-m_1)}{m^5_0m^5_1M^4}\frac{L^5_{1l}}{L^3_{2l}G^5_{2l}}
\sqrt{1-\frac{G^2_{2l}}{L^2_{2l}}}\frac{G_{1l}}{\sqrt{L^2_{1l}-G^2_{1l}}}(13-\nonumber\\
& & -9\frac{G^2_{1l}}{L^2_{1l}})\cos{(g_{1l}-g_{2l})}\label{geq} 
\end{eqnarray}

If we replace the Delaunay variables from equations (\ref{eq:delaunay}) with the corresponding orbital elements and use 
${e_{21l}=e_{2l}\cos{g_{2l}}}$ and ${e_{22l}=e_{2l}\sin{g_{2l}}}$, the above system becomes:

\begin{eqnarray}
\frac{de_{21l}}{dt} & = & -\frac{3}{8}\frac{\sqrt{\mathcal{G}M}m_0m_1a^2_{1l}}{(m_0+m_1)^2a^{\frac{7}{2}}_{2l}(1-e^2_{2l})^2}(2+3e^2_{1l})e_{22l}+\nonumber\\
& & + \frac{15}{64}\frac{\sqrt{\mathcal{G}M}m_0m_1(m_0-m_1)a^3_{1l}}{(m_0+m_1)^3a^{\frac{9}{2}}_{2l}(1-e^2_{2l})^3}e_{1l}(4+3e^2_{1l})\big[(1+4e^2_{22l}-e^2_{21l})\sin{g_{1l}}+\nonumber\\
& & +5e_{21l}e_{22l}\cos{g_{1l}}\big]\\
\frac{de_{22l}}{dt} & = & \frac{3}{8}\frac{\sqrt{\mathcal{G}M}m_0m_1a^2_{1l}}{(m_0+m_1)^2a^{\frac{7}{2}}_{2l}(1-e^2_{2l})^2}(2+3e^2_{1l})e_{21l}+\nonumber\\
& & + \frac{15}{64}\frac{\sqrt{\mathcal{G}M}m_0m_1(m_0-m_1)a^3_{1l}}{(m_0+m_1)^3a^{\frac{9}{2}}_{2l}(1-e^2_{2l})^3}e_{1l}(4+3e^2_{1l})\big[(-1+e^2_{22l}-4e^2_{21l})\cos{g_{1l}}-\nonumber\\
& & -5e_{21l}e_{22l}\sin{g_{1l}}\big]\\
\label{post}
\frac{dg_{1l}}{dt} & = &\frac{3}{4}\frac{\sqrt{\mathcal{G}}m_2a^{\frac{3}{2}}_{1l}\sqrt{1-e^2_{1l}}}{(m_0+m_1)^{\frac{1}{2}}a^3_{2l}(1-e^2_{2l})^{\frac{3}{2}}}-\frac{15}{64}\frac{\sqrt{\mathcal{G}}m_2(m_0-m_1)a^{\frac{5}{2}}_{1l}e_{2l}\sqrt{1-e^2_{1l}}}{(m_0+m_1)^{\frac{3}{2}}a^4_{2l}e_{1l}\sqrt{1-e^2_{2l}}}(4+\nonumber\\
& & +9e^2_{1l})\cos(g_{1l}-g_{2l}).
\end{eqnarray}

Since the planetary orbit is initially circular and its eccentricity is expected to remain small,
we neglect terms of order ${O(e^2_{2l})}$ in the above equations and we also keep the dominant term in the equation for the  stellar pericentre. As a result, the equations of motion
assume the following form:
\begin{eqnarray}
\frac{de_{21l}}{dt} & = & -K_1e_{22l}+K_2\sin{g_{1l}},\\
\frac{de_{22l}}{dt} & = & K_1e_{21l}-K_2\cos{g_{1l}},\\
\frac{dg_{1l}}{dt} & = & K_3,
\end{eqnarray}
where
\begin{eqnarray}
\begin{split}
K_1 & = \frac{3}{8}\frac{\sqrt{\mathcal{G}M}m_0m_1a^2_{1l}}{(m_0+m_1)^2a^{\frac{7}{2}}_{2l}}(2+3e^2_{1l})\\
K_2 & = \frac{15}{64}\frac{\sqrt{\mathcal{G}M}m_0m_1(m_0-m_1)a^3_{1l}}{(m_0+m_1)^3a^{\frac{9}{2}}_{2l}}e_{1l}(4+3e^2_{1l})\\
K_3 & = \frac{3}{4}\frac{\sqrt{\mathcal{G}}m_2a^{\frac{3}{2}}_{1l}\sqrt{1-e^2_{1l}}}{(m_0+m_1)^{\frac{1}{2}}a^3_{2l}} \label{eq:K}
\end{split}
\end{eqnarray}
are constants.

The above system of differential equations can be solved analytically and the solution is
\begin{eqnarray}
\begin{split}
e_{21l}(t) & = & C_{1}\cos{K_1t}+C_{2}\sin{K_1t}+\frac{K_2}{K_1-K_3}\cos{K_3t}\label{eq:e2l}\\
e_{22l}(t) & = & C_{1}\sin{K_1t}-C_{2}\cos{K_1t}+\frac{K_2}{K_1-K_3}\sin{K_3t},
\end{split}
\end{eqnarray} 
for the planetary orbit, where $C_{1}$ and $C_{2}$ are constants of integration.  For the stellar orbit we get 
\begin{equation}
g_{1l}=K_3t,
\end{equation}
where we have assumed without loss of generality that the initial longitude of pericenter of the inner orbit is zero.
Consequently, the components of the eccentric vector of the stellar orbit become
\begin{eqnarray}
\begin{split}
e_{11l}(t) & = & e_1 \cos{g_{1l}} = e_{1} \cos{(K_3 t)} \label{eq:e1l}\\
e_{12l}(t) & = & e_1 \sin{g_{1l}} = e_{1} \sin{(K_3 t)},
\end{split}
\end{eqnarray}

So far our results are in agreement with other authors, e.g. \citet{Moriwaki-2004},
\citet{2014arXiv1407.5493D}.

\subsection{Calculation of the short and mid term evolution}
\begin{figure}
\begin{center}
\includegraphics[width=90mm]{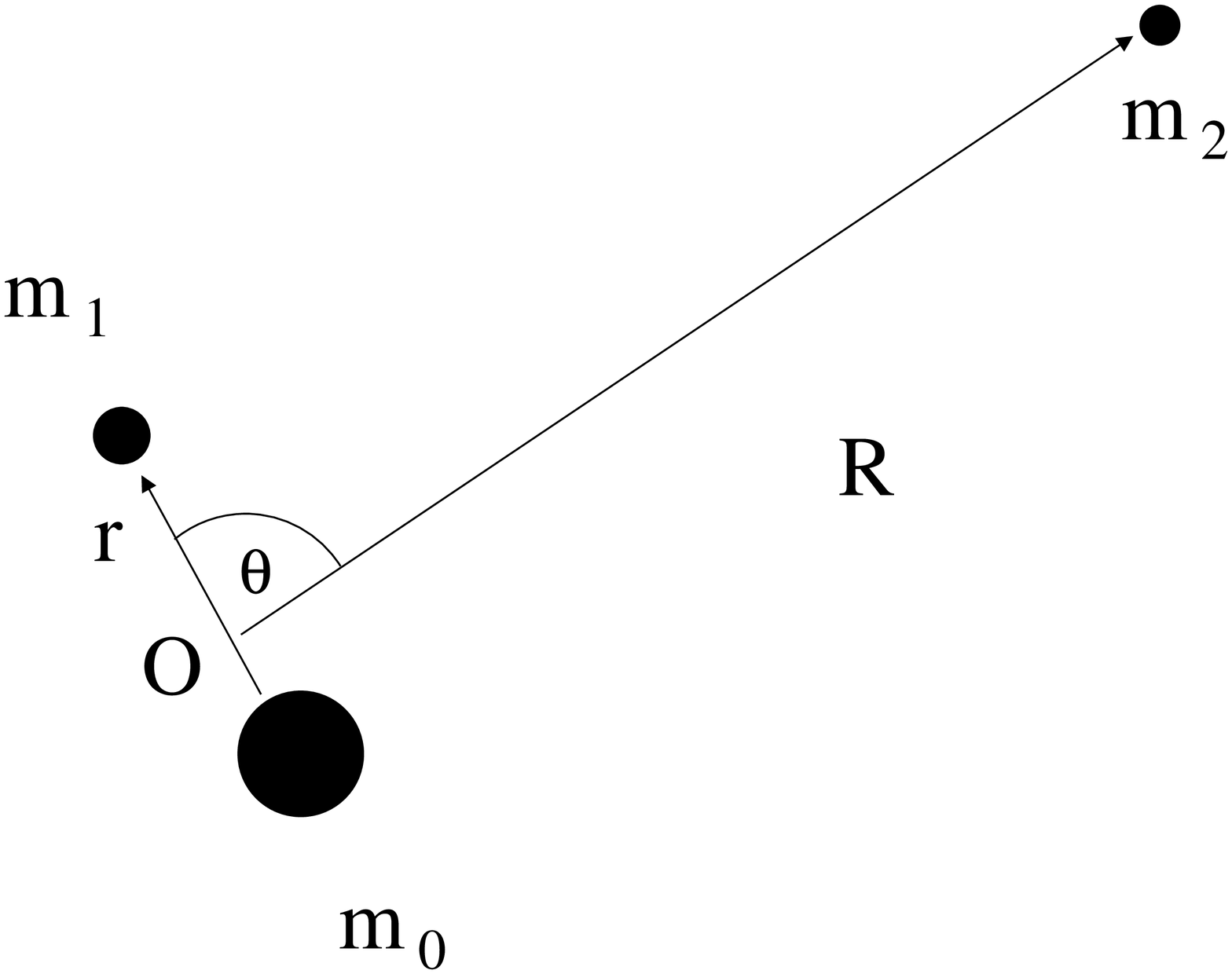}
\caption[]{The Jacobi decomposition of the three body problem.\label{fig:jacobi}}
\label{fig4}
\end{center}
\end{figure}
Having the long term solution for the eccentricity evolution of our system at our disposal, we now proceed to derive equations for
the mid and short periodic terms.
Once again, we use the Jacobi
decomposition of the three body problem to describe the motion of the system (see Figure \ref{fig:jacobi}). In that context,
the equation of motion of the outer body is: 
\begin{equation}
\ddot{\textit{\textbf{R}}}=-\mathcal{G}M\Big(\mu_{0}\frac{\textit{\textbf{R}}+\mu_{1}\textit{\textbf{r}}}{|\textit{\textbf{R}}+\mu_{1}\textit{\textbf{r}}|^3}+\mu_{1}\frac{\textit{\textbf{R}}-\mu_{0}\textit{\textbf{r}}}{|\textit{\textbf{R}}-\mu_{0}\textit{\textbf{r}}|^3}\Big)=\mathcal{G}M\frac{\partial}{\partial{\textit{\textbf{R}}}}\Big(\frac{\mu_{0}}{|\textit{\textbf{R}}+\mu_{1}\textit{\textbf{r}}|}+\frac{\mu_1}{|\textit{\textbf{R}}-\mu_{0}\textit{\textbf{r}}|}\Big)
\label{eqmo}
\end{equation}
with 
\begin{displaymath}
\mu_i=\frac{m_i}{m_0+m_1},\qquad i=0,1. 
\end{displaymath}

Once more, since we work with a hierarchical triple system and the third body is assumed to be at considerable distance from
the inner binary, implying that 
\begin{math}
r/R 
\end{math}
is small, the inverse distances in equation (\ref{eqmo}) can be expressed in terms of Legendre polynomials as
\begin{eqnarray}
\frac{1}{|\textit{\textbf{R}}+\mu_{1}\textit{\textbf{r}}|}&=&\frac{1}{R}\sum^{\infty}_{n=0}
\left(- \frac{\mu_{1}r}{R} \right)^{n}\mathcal{P}_{{\rm n}}(\cos{\theta}),\\
\mbox{and}\nonumber\\
\frac{1}{|\textit{\textbf{R}}-\mu_{0}\textit{\textbf{r}}|}&=&\frac{1}{R}\sum^{\infty}_{n=0}
\left( \frac{\mu_{0}r}{R} \right) ^{n}\mathcal{P}_{{\rm n}}(\cos{\theta}).
\end{eqnarray}
Expanding to second
order and making the substitution 
\begin{displaymath}
\cos{\theta}=\frac{\textit{\textbf{r}}\cdot\textit{\textbf{R}}}{rR},
\end{displaymath}
the equation of motion of the outer binary becomes
\begin{equation}
\ddot{\textit{\textbf{R}}}=\mathcal{G}M\Big\{-\frac{\textit{\textbf{R}}}{R^3}+\mu_0\mu_1\big[3\frac{(\textit{\textbf{r}}\cdot\textit{\textbf{R}})\textit{\textbf{r}}}{R^5}-\frac{15}{2}
\frac{(\textit{\textbf{r}}\cdot\textit{\textbf{R}})^2\textit{\textbf{R}}}{R^7}+\frac{3}{2}\frac{r^2\textit{\textbf{R}}}{R^5}\big]\Big\}
\label{eqmo2}
\end{equation}
Using the Runge-Lenz vector of the outer orbit we can obtain an expression for the eccentricity.  Hence:

\begin{eqnarray}
\label{eqev}
\textit{\textbf{e}}_{2}&=&-\frac{\textit{\textbf{R}}}{R}+\frac{1}{\mathcal{G}M}(\dot{\textit{\textbf{R}}}\times\textit{\textbf{h}})\Rightarrow\nonumber\\
\dot{\textit{\textbf{e}}}_{2}&=&-\frac{\dot{\textit{\textbf{R}}}}{R}+\frac{\dot{R}}{R^2}\textit{\textbf{R}}+\frac{1}{\mathcal{G}M}[2(\dot{\textit{\textbf{R}}}\cdot\ddot{\textit{\textbf{R}}})\textit{\textbf{R}}-(\textit{\textbf{R}}\cdot\ddot{\textit{\textbf{R}}})\dot{\textit{\textbf{R}}}-
 (\textit{\textbf{R}}\cdot\dot{\textit{\textbf{R}}})\ddot{\textit{\textbf{R}}}],
\end{eqnarray}
where ${\textit{\textbf{h}}=\textit{\textbf{R}}\times\dot{\textit{\textbf{R}}}}$.

Substituting equation (\ref{eqmo}) into equation (\ref{eqev}) and assuming that the eccentricity remains small (and hence neglecting the
term $\textit{\textbf{R}}\cdot\dot{\textit{\textbf{R}}}$), we obtain:
\begin{equation}
\dot{\textit{\textbf{e}}}_{2sm}=\frac{m_0m_1}{(m_0+m_1)^2}\big[6\frac{(\textit{\textbf{r}}\cdot\textit{\textbf{R}})(\textit{\textbf{r}}\cdot\dot{\textit{\textbf{R}})}}{R^5} \textit{\textbf{R}}+\frac{9}{2}\frac{(\textit{\textbf{r}}\cdot\textit{\textbf{R}})^2}{R^5}\dot{\textit{\textbf{R}}}-\frac{3}{2}\frac{r^2}{R^3}\dot{\textit{\textbf{R}}}\big],
\label{eqev2}
\end{equation}
where the indices $s$ and $m$ refer to short and medium period terms as explained earlier.

Now, the Jacobi vectors can be represented 
as ${\textit{\textbf{r}}=a_1(\cos{E_1}-e_1,\sqrt{1-e^2_1}\sin{E_1})}$
and ${\textit{\textbf{R}}=a_{2}(\cos{l_2},\sin{l_2})}$, where ${a_1}$, ${e_1}$, ${E_1}$, are the semi-major axis, 
the eccentricity and the eccentric anomaly of the inner orbit, while ${a_2}$ and ${l_2}$ are the semi major-axis and the mean anomaly 
of the outer orbit respectively. 
Substituting for ${\textit{\textbf{r}}}$ and ${\textit{\textbf{R}}}$ in equation (\ref{eqev2}), we acquire for the rate of change of the componets of the outer eccentric vector:

\begin{eqnarray}
\begin{split}
\dot{e}_{21sm}(t) & =  \frac{m_{0}m_{1}}{(m_{0}+m_{1})^{\frac{4}{3}}M^{\frac{2}{3}}}\frac{n_2}{X^{\frac{4}{3}}}\Big[6\big[(\cos{E_1}-e_1)\cos{l_2}+\sqrt{1-e^2_1}\sin{E_1}\sin{l_2}\big]\times\\
& \times\big[-(\cos{E_1}-e_1)\sin{l_2}+\sqrt{1-e^2_1}\sin{E_1}\cos{l_2}\big]\cos{l_2}-\frac{9}{2}\big[(\cos{E_1}-\\
& -e_1)\cos{l_2}+\sqrt{1-e^2_1}\sin{E_1}\sin{l_2}\big]^2\sin{l_2}+\frac{3}{2}(1-e_1\cos{E_1})^2\sin{l_2}\Big] \label{eqr1}\\
\dot{e}_{22sm}(t) & =  \frac{m_{0}m_{1}}{(m_{0}+m_{1})^{\frac{4}{3}}M^{\frac{2}{3}}}\frac{n_2}{X^{\frac{4}{3}}}\Big[6\big[(\cos{E_1}-e_1)\cos{l_2}+\sqrt{1-e^2_1}\sin{E_1}\sin{l_2}\big]\times\\
& \times\big[-(\cos{E_1}-e_1)\sin{l_2}+\sqrt{1-e^2_1}\sin{E_1}\cos{l_2}\big]\sin{l_2}+\frac{9}{2}\big[(\cos{E_1}-\\
& -e_1)\cos{l_2}+\sqrt{1-e^2_1}\sin{E_1}\sin{l_2}\big]^2\cos{l_2}-\frac{3}{2}(1-e_1\cos{E_1})^2\cos{l_2}\Big],
\end{split}
\end{eqnarray}
where X is the period ratio of the two orbits.

By using equations (\ref{eqr1}), we are now able to obtain the short period terms for the evolution of the planetary eccentric vector.  This will be done in two steps.  
First, we average the above equations over the fast motion and then we  integrate over time:
\begin{eqnarray}
\begin{split}
\label{eq:r11}
e_{21m}(t) & = & \frac{1}{16}\frac{m_{0}m_{1}}{(m_{0}+m_{1})^{\frac{4}{3}}M^{\frac{2}{3}}}\frac{1}{X^{\frac{4}{3}}}\Big[12\cos{l_2}+e^2_1\big(33\cos{l_2}+35\cos{3l_2}\big)\Big]+C_{e_{21m}}\\
e_{22m}(t) & = & \frac{1}{16}\frac{m_{0}m_{1}}{(m_{0}+m_{1})^{\frac{4}{3}}M^{\frac{2}{3}}}\frac{1}{X^{\frac{4}{3}}}\Big[12\sin{l_2}+e^2_1\big(3\sin{l_2}+35\sin{3l_2}\big)\Big]+C_{e_{22m}}.
\end{split}
\end{eqnarray}
Here, ${C_{e_{21m}}}$ and ${C_{e_{22m}}}$ are constants of integration.
In a second step we add higher frequency terms by integrating equations (\ref{eqr1}) with respect to time.  During that process, any terms that appear and have
already been taken into consideration (given by equations (\ref{eq:r11})) are removed. Integrating by
parts yields a power series solution in terms of ${1/X}$.
By keeping the largest term in that power series we finally end up with the following short term equations for the eccentric vector of the planetary orbit:
\begin{eqnarray}
\begin{split}
\label{eq:s11}
e_{21s}(t) & = & \frac{m_{0}m_{1}}{(m_{0}+m_{1})^{\frac{4}{3}}M^{\frac{2}{3}}}\frac{1}{X^{\frac{7}{3}}} P_1(t)+C_{e_{21s}}\\
e_{22s}(t) & = & \frac{m_{0}m_{1}}{(m_{0}+m_{1})^{\frac{4}{3}}M^{\frac{2}{3}}}\frac{1}{X^{\frac{7}{3}}} P_2(t)+C_{e_{22s}},
\end{split}
\end{eqnarray}
with ${P_1(t)}$ and ${P_2(t)}$ given in the Appendix. 
Combining equations (\ref{eq:r11}) and (\ref{eq:s11}) we find formulae for all non-secular contributions: 
\begin{eqnarray}
\begin{split}
\label{eq:sm11}
e_{21sm}(t) & = & \frac{m_{0}m_{1}}{(m_{0}+m_{1})^{\frac{4}{3}}M^{\frac{2}{3}}}\frac{1}{X^{\frac{4}{3}}}\Big[\frac{3}{4}\cos{l_2}+e^2_1\big(\frac{33}{16}\cos{l_2}+\frac{35}{16}\cos{3l_2}\big)+\frac{P_1(t)}{X}\Big]+C_{e_{21sm}}\\
e_{22sm}(t) & = & \frac{m_{0}m_{1}}{(m_{0}+m_{1})^{\frac{4}{3}}M^{\frac{2}{3}}}\frac{1}{X^{\frac{4}{3}}}\Big[\frac{3}{4}\sin{l_2}+e^2_1\big(\frac{3}{16}\sin{l_2}+\frac{35}{16}\sin{3l_2}\big)+\frac{P_2(t)}{X}\Big]+C_{e_{22sm}}.
\end{split}
\end{eqnarray}
 We would like to point out here that $a_1$, $a_2$ and $e_1$ have been treated as constants in the above calculations.

\subsection{Complete solution of the eccentricities}
We can now combine equations (\ref{eq:e2l}) and (\ref{eq:sm11}) in order
to obtain the total solution for the outer eccentricity. As in the case of the 
inner eccentricity, this is done by substituting the constants of integration in
the short period solution with the expressions for the secular evolution:  
 \begin{eqnarray}
 \begin{split}
 e_{21}(t) & = & e_{21s}(t)+e_{21m}(t)+e_{21l}(t) -C_{e_{21sm}}\label{f1}\\
 e_{22}(t) & = & e_{22s}(t)+e_{22m}(t)+e_{22l}(t) -C_{e_{22sm}}.
 \end{split}
 \end{eqnarray}
The constants in equation (\ref{f1}) are determined by the fact that we assume the outer eccentricity to be initially zero, i.e.
\begin{equation}
e_{21}(t_0) = e_{22}(t_0) = 0
\end{equation}
which eventually leads to
\begin{eqnarray}
\begin{split}
e_{21}(t)  = & \frac{m_{0}m_{1}}{(m_{0}+m_{1})^{\frac{4}{3}}M^{\frac{2}{3}}}\frac{1}{X^{\frac{4}{3}}}\Big[\frac{3}{4}\cos{l_2}+e^2_1\big(\frac{33}{16}\cos{l_2}+\frac{35}{16}\cos{3l_2}\big)+\frac{P_1(t)}{X}\Big]+\\ 
             & C_{1}\cos{K_1t}+C_{2}\sin{K_1t}+\frac{K_2}{K_1-K_3}\cos{K_3t}\\
e_{22}(t)  = & \frac{m_{0}m_{1}}{(m_{0}+m_{1})^{\frac{4}{3}}M^{\frac{2}{3}}}\frac{1}{X^{\frac{4}{3}}}\Big[\frac{3}{4}\sin{l_2}+e^2_1\big(\frac{3}{16}\sin{l_2}+\frac{35}{16}\sin{3l_2}\big)+\frac{P_2(t)}{X}\Big]+ \\
             & C_{1}\sin{K_1t}-C_{2}\cos{K_1t}+\frac{K_2}{K_1-K_3}\sin{K_3t}\label{eq:f2},
\end{split}
\end{eqnarray}
where 
\begin{eqnarray}
C_{1} & = & -\frac{m_{0}m_{1}}{(m_{0}+m_{1})^{\frac{4}{3}}M^{\frac{2}{3}}}\frac{1}{X^{\frac{4}{3}}}\Big[\frac{3}{4}\cos{l_{20}}+e^2_1\big(\frac{33}{16}\cos{l_{20}}+\frac{35}{16}\cos{3l_{20}}\big)+\nonumber\\
& & +\frac{P_1(t_0)}{X}\Big]-\frac{K_2}{K_1-K_3}\\
C_{2} & = & \frac{m_{0}m_{1}}{(m_{0}+m_{1})^{\frac{4}{3}}M^{\frac{2}{3}}}\frac{1}{X^{\frac{4}{3}}}\Big[\frac{3}{4}\sin{l_{20}}+e^2_1\big(\frac{3}{16}\sin{l_{20}}+\frac{35}{16}\sin{3l_{20}}\big)+\nonumber\\
& & +\frac{P_2(t_0)}{X}\Big].
\end{eqnarray}
The quantities $K_i$ have been defined in equations (\ref{eq:K}) and ${l_{20}}$ is the initial planetary mean anomaly.
The evolution of the eccentricity of the inner orbit is given by equations (\ref{eq:e1l}).

\subsection{Maximum and average squared outer eccentricity}
We can also obtain an estimate for the maximum value the outer eccentricity is bound to reach during its evolution assuming it was initially zero.
This is achieved by doing some algebraic manipulations and by maximizing trigonometric functions in equations (\ref{eq:e2l}) and (\ref{eq:sm11}).
Eventually we obtain:
\begin{eqnarray}
\begin{split}
e^{max}_{2}= & \quad e^{max}_{2sm}+e^{max}_{2l} \\
 = & \quad \frac{m_{0}m_{1}}{(m_{0}+m_{1})^{\frac{4}{3}}M^{\frac{2}{3}}}\frac{1}{X^{\frac{4}{3}}}\bigg[\frac{3}{2}+
\frac{17}{2}e^2_1+\frac{1}{X}\Big(3+19e_1+\frac{21}{8}e^2_1-\frac{3}{2}e^3_1\Big)\bigg] \\
 & \quad + \frac{2K_2}{K_1-K_3}\label{eq:e2max}.
\end{split}
\end{eqnarray}

Averaging equations (\ref{eq:f2}) over time we obtain the following expression (excluding the short period terms):
\begin{eqnarray}
\langle e^2_2 \rangle_t & = & \langle e^2_{21}(t)+e^2_{22}(t)\rangle_t=\frac{m^2_{0}m^2_{1}}{(m_{0}+m_{1})^{\frac{8}{3}}M^{\frac{4}{3}}}\frac{1}{X^{\frac{8}{3}}}
\bigg[\frac{9}{8}+\frac{27}{8}e^2_1+\frac{887}{64}e^4_1+\nonumber\\
& & +\frac{75}{16}e^2_1+e^4_1\Big(\frac{225}{32}\cos{2l_{20}}+\frac{525}{128}\cos{4l_{20}}\Big)\bigg]+\nonumber\\
& & +\frac{m_{0}m_{1}}{(m_{0}+m_{1})^{\frac{4}{3}}M^{\frac{2}{3}}}\frac{1}{X^{\frac{4}{3}}}\frac{K_{2}}{K_1-K_3}\bigg[
\frac{3}{2}\cos{l_{20}}+e^2_1\Big(\frac{33}{8}\cos{l_{20}}+\nonumber\\
& & +\frac{35}{8}\cos{3l_{20}}\Big)\Bigg]+2\Big(\frac{K_{2}}{K_1-K_3}\Big)^2.
\end{eqnarray}

If equations (\ref{eq:f2}) are averaged over both time and initial angles, we find the most compact expression for
the averaged square outer eccentricity:
\begin{eqnarray}
\label{eq:e2av}
\langle e^2_2\rangle & = & \langle e^2_{21}(t)+e^2_{22}(t)\rangle=\frac{m^2_{0}m^2_{1}}{(m_{0}+m_{1})^{\frac{8}{3}}M^{\frac{4}{3}}}\frac{1}{X^{\frac{8}{3}}}
\bigg[\frac{9}{8}+\frac{27}{8}e^2_1+\frac{887}{64}e^4_1-\nonumber\\
& & -\frac{975}{64}\frac{1}{X}e^4_1\sqrt{1-e^2_1}+\frac{1}{X^2}\Big(\frac{225}{64}+\frac{6619}{64}e^2_1-\frac{26309}{512}e^4_1-\frac{393}{64}e^6_1\Big)\bigg]+\nonumber\\
& & +2\Big(\frac{K_{2}}{K_1-K_3}\Big)^2.
\end{eqnarray}

\section{NUMERICAL TESTING}
\label{sec:test}
In order to test the validity of our analytical estimates,  we integrated the full equations of motion numerically.  For that purpose, we used the symplectic integrator with time transformation developed by \citet{1997CeMDA..67..145M}, 
specially designed to integrate hierarchical triple systems.  The code uses standard Jacobi coordinates, i.e. it calculates the relative position and velocity vectors of the inner and outer orbit at every time step. 
Those were used to generate orbital elements of the stellar and planetary orbits.  

We investigated systems with nine different mass combinations. Stellar masses were chosen such as $m_1/(m_0+m_1)=0.5, 0.3, 0.1$.
The planetary masses we used were such that $m_2/(m_0+m_1)=10^{-2}, 10^{-3}, 10^{-6}$, i.e. we chose Earth-like, Jovian and brown dwarf like companions.  The eccentricity ranged from 0 to 0.9, while for the semimajor axis the range was between one and 12 times the critical
semi-major axis as given in \citet{1999AJ....117..621H}, who studied the stability of planetary orbits in binary star systems by performing numerical simulations in the framework of the planar elliptic restricted three body problem.  If a particle survived the whole integration time at all initial longitudes, then the system was classified as stable.  For circumbinary orbits, and by using a least squares fit to their data, they obtained an expression for the critical planetary semimajor axis $a_c$, i.e.  the smallest planetary semi-major axis for which the particles were stable at all initial positions, given by (using our notation for the mass parameters and orbital elements):
\begin{eqnarray}
a_{c} & = & [(1.60\pm 0.04)+(5.10\pm 0.05)e_1+(-2.22\pm 0.11)e_1^{2}+(4.12\pm 0.09)\mu_1+\nonumber\\
& & +(-4.27\pm 0.17)e_1\mu_1+(-5.09 \pm 0.11)\mu_1^{2}+(4.61\pm 0.36)e_1^{2}\mu_1^{2}]a_1.
\label{hol2}
\end{eqnarray}

Finally, the initial stellar eccentric anomaly and the initial planetary mean anomaly were varied from $0^{\circ}$ to $360^{\circ}$ 
with a step of $45^{\circ}$.  The results from the numerical simulations were compared with our analytical estimates for $\langle e^2_2\rangle$
and $e^{max}_2$, i.e. equations (\ref{eq:e2max}) and (\ref{eq:e2av}), on both short and secular timescales.  For the short
timescale simulations the integration time was one planetary orbital period, while for the long 
term simulations the integration time was set to $2\pi/(K_1-K_3)$, which is the analytical 
estimate for the secular period of the planetary eccentricity.

\begin{figure}
\begin{center}
\includegraphics[width=80mm]{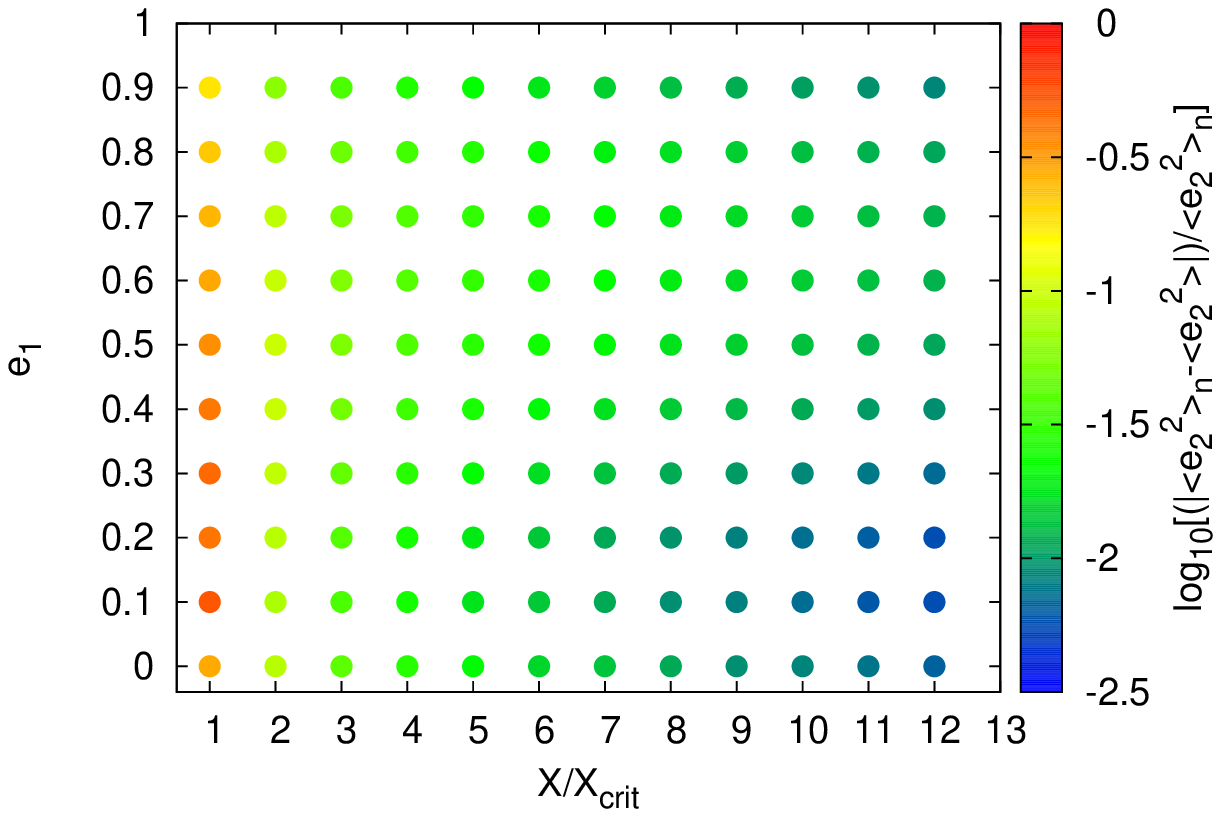}
\includegraphics[width=80mm]{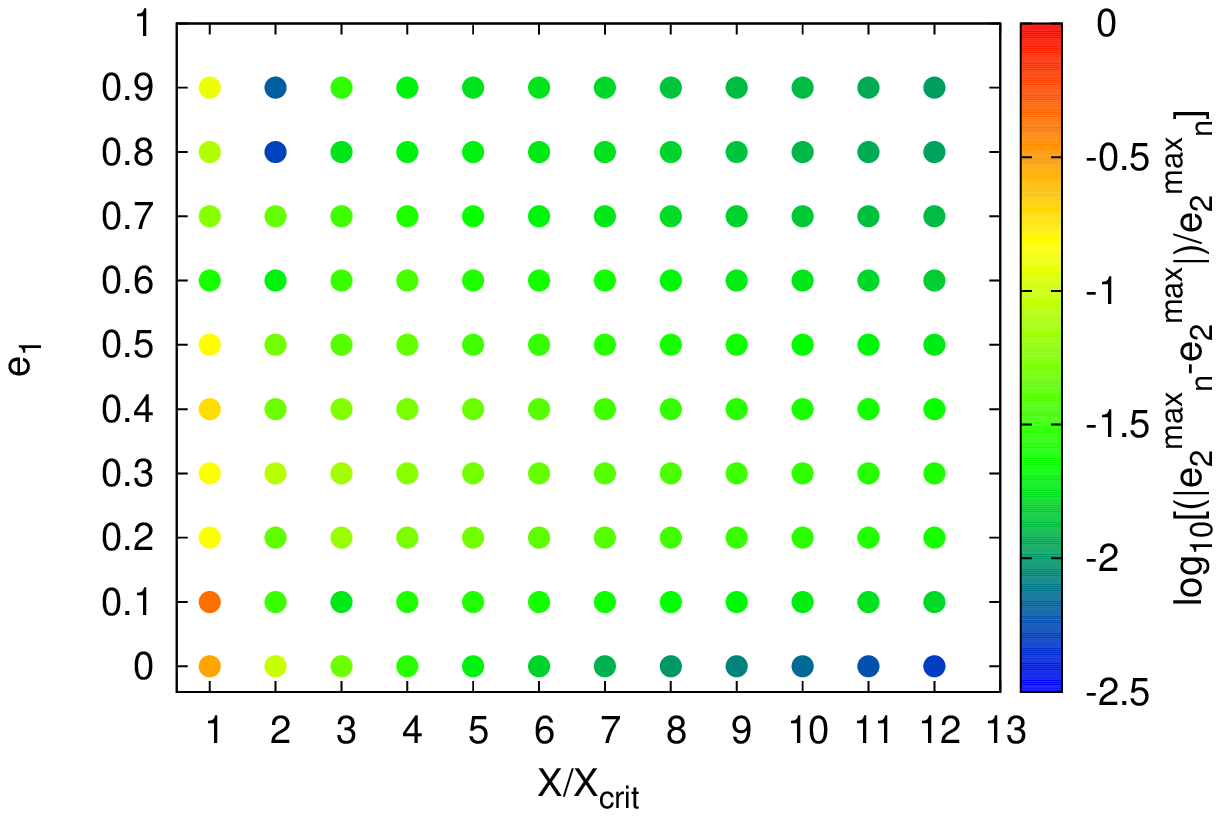} 
\includegraphics[width=80mm]{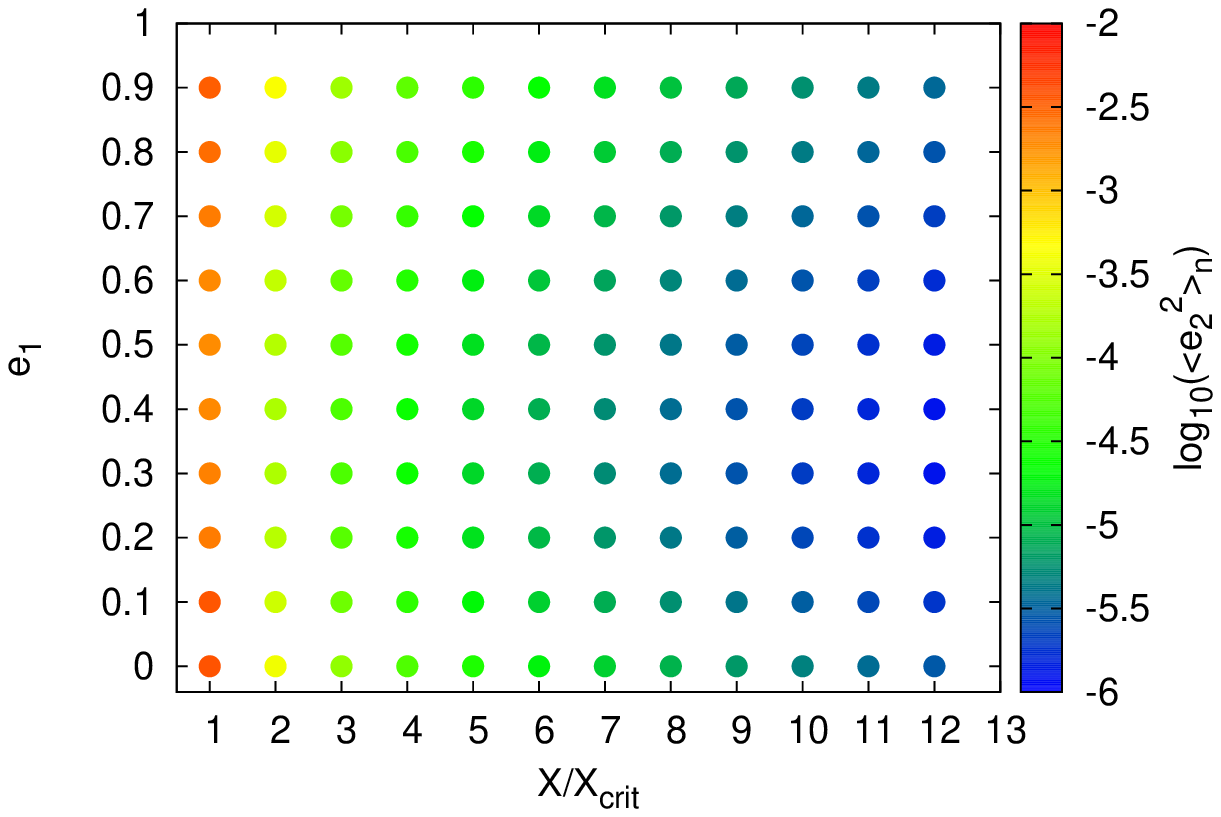}
\includegraphics[width=80mm]{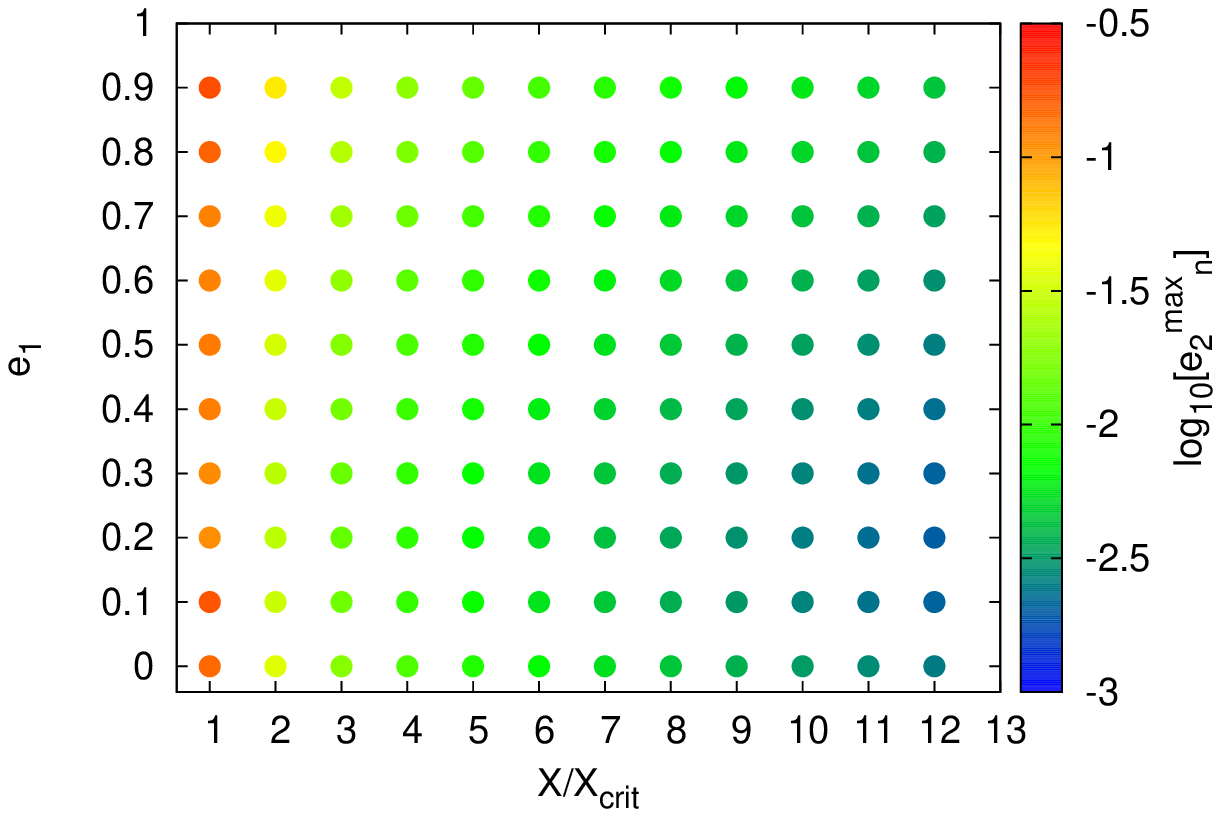} 

\caption[]{ Top row: logarithmic error (in color) for the averaged square planetary eccentricity (left) and the maximum planetary eccentricity (right). Bottom row: the logarithms of the numerical values (in colour) of the averaged squared planetary eccentricity (left) and the maximum planetary eccentricity (right).  The mass parameters of the system are $m_1/(m_0+m_1)=0.5$ and $m_2/(m_0+m_1)=10^{-2}$, $X$ is the initial period ratio and $X_{crit}$ is the critical period ratio based on \citet{1999AJ....117..621H}.  The integration time is one planetary orbital period.
\label{fig:test1}}
\end{center}
\end{figure}

\begin{figure}
\begin{center}
\includegraphics[width=80mm]{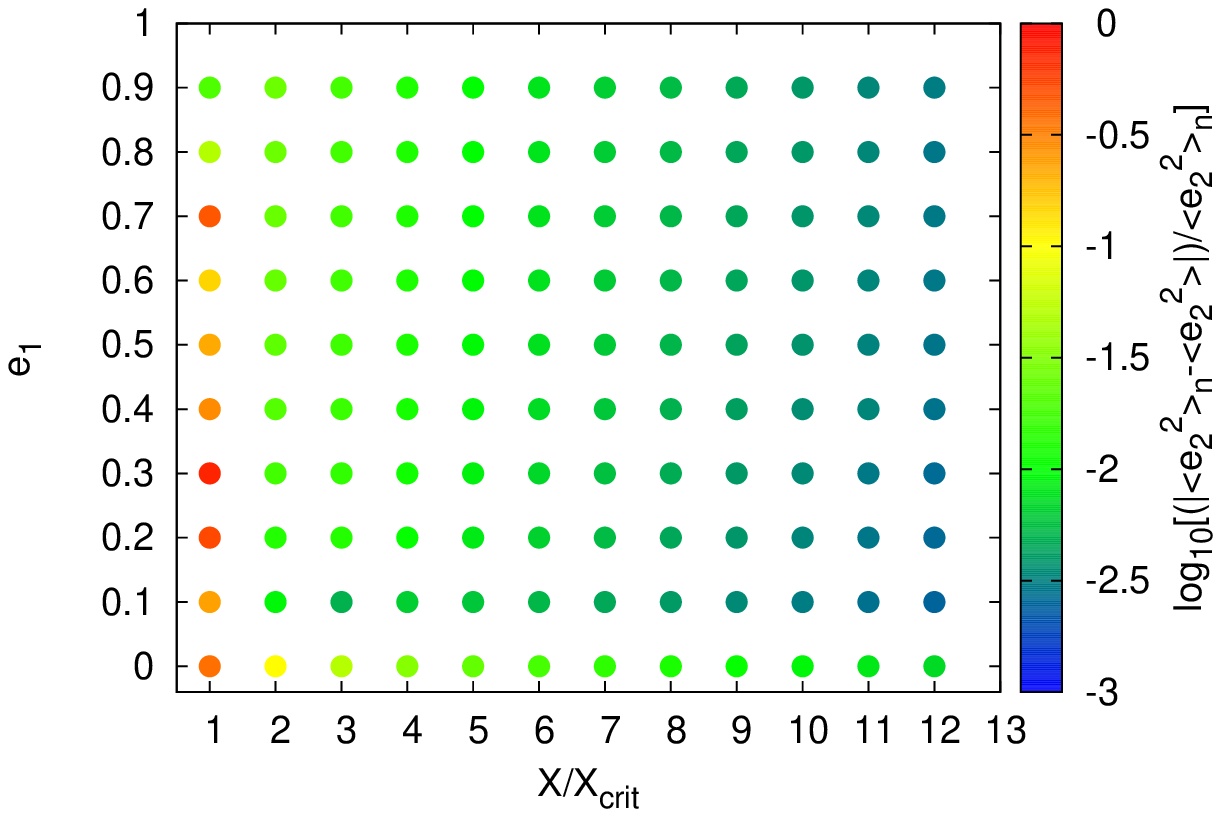}
\includegraphics[width=80mm]{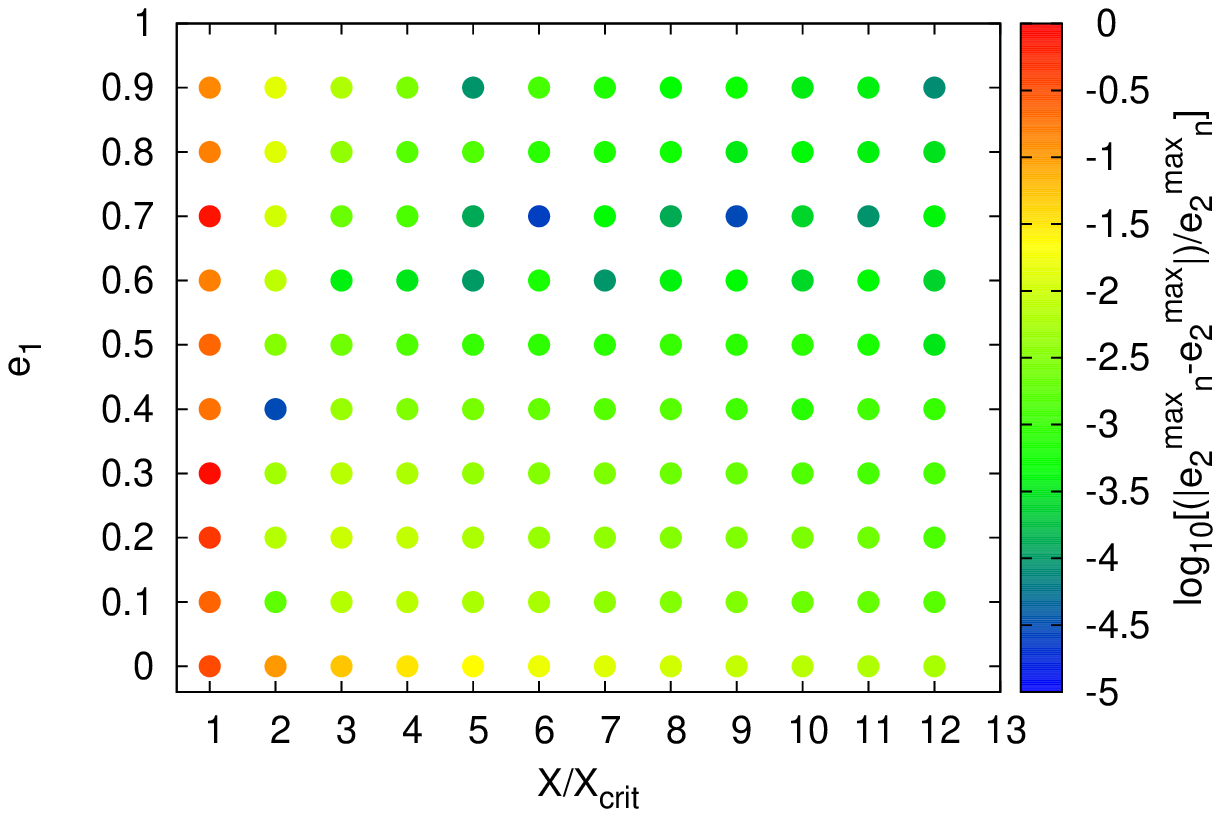}
\includegraphics[width=80mm]{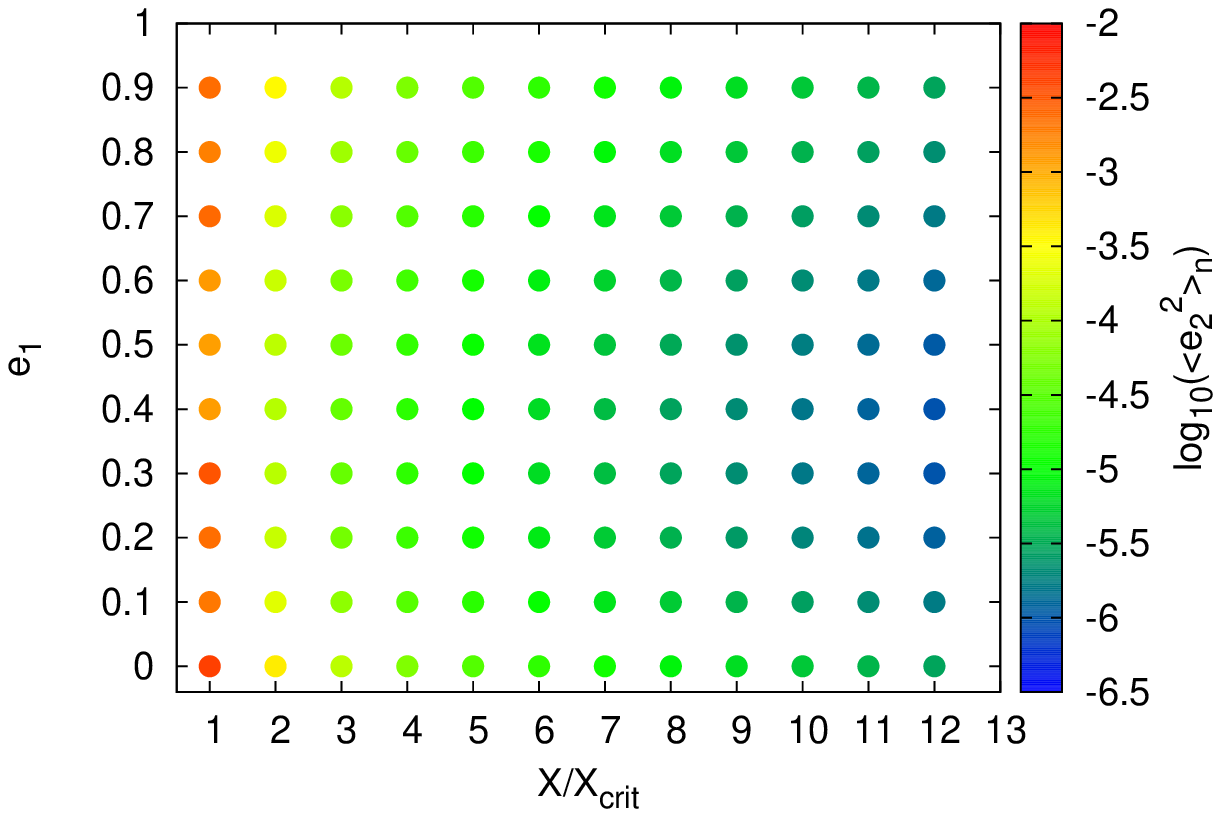}
\includegraphics[width=80mm]{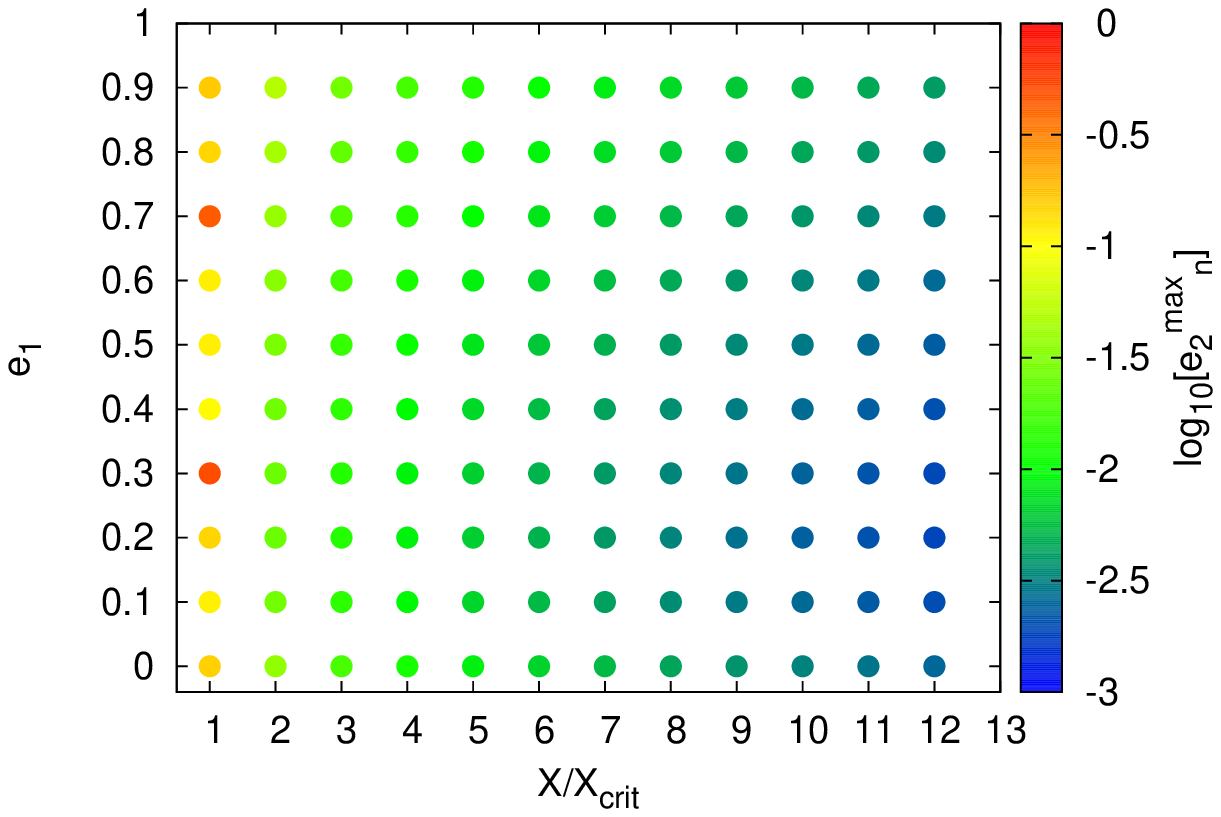}

\caption[]{ Top row: logarithmic error (in color) for the averaged square planetary eccentricity (left) and the maximum planetary eccentricity (right).  Bottom row: the logarithms of the numerical values (in colour) of the averaged squared planetary eccentricity (left) and the maximum planetary eccentricity (right).  The mass parameters of the system are $m_1/(m_0+m_1)=0.3$ and $m_2/(m_0+m_1)=10^{-3}$, $X$ is the initial period ratio and $X_{crit}$ is the critical period ratio based on \citet{1999AJ....117..621H}.  The integration time is one analytical secular period.  
\label{fig:test2}}
\end{center}
\end{figure}

\begin{figure}
\begin{center}
\includegraphics[width=80mm]{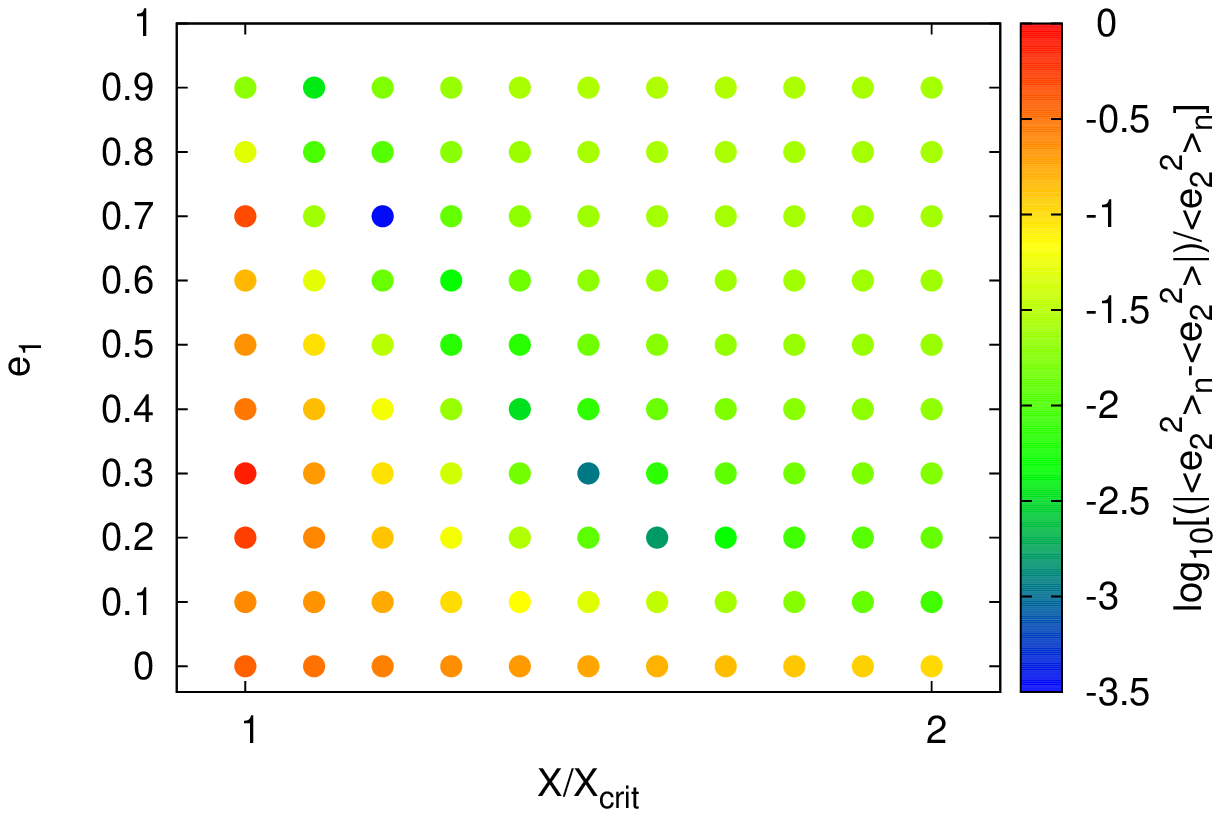}
\includegraphics[width=80mm]{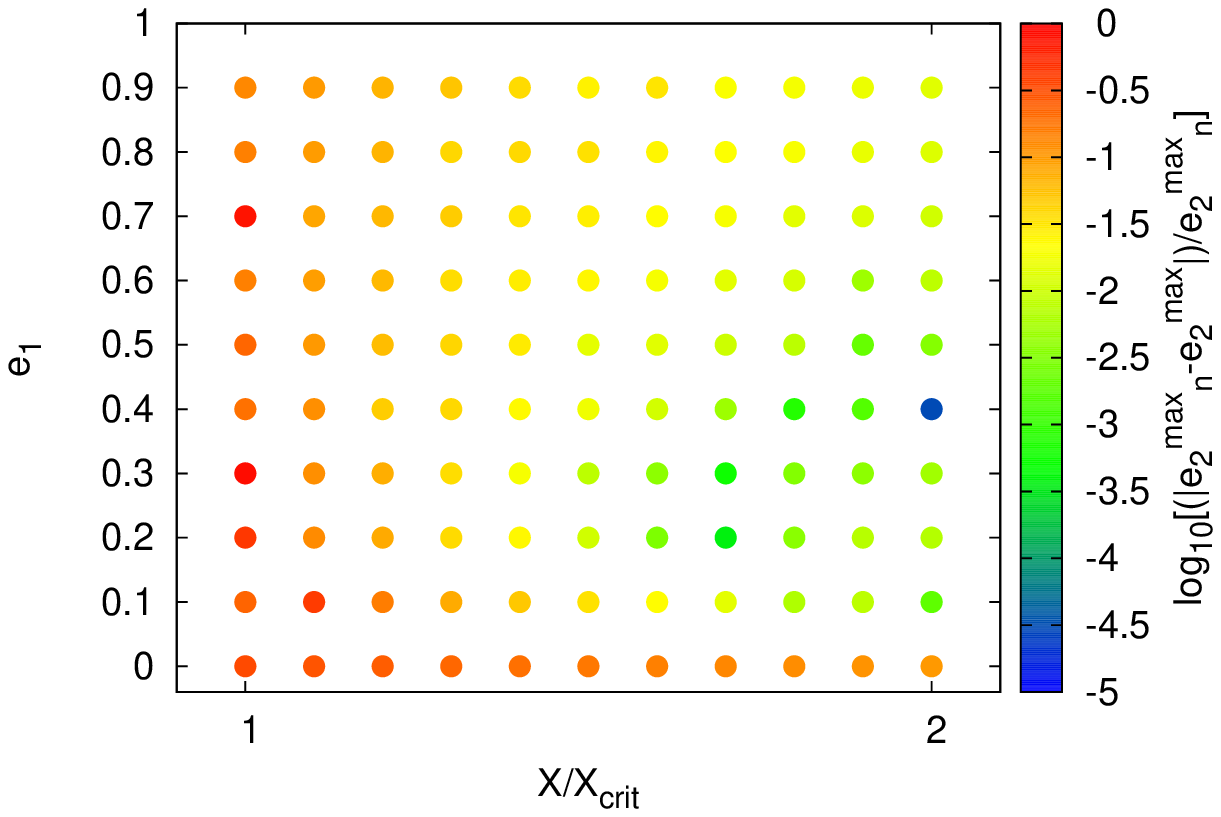}
\includegraphics[width=80mm]{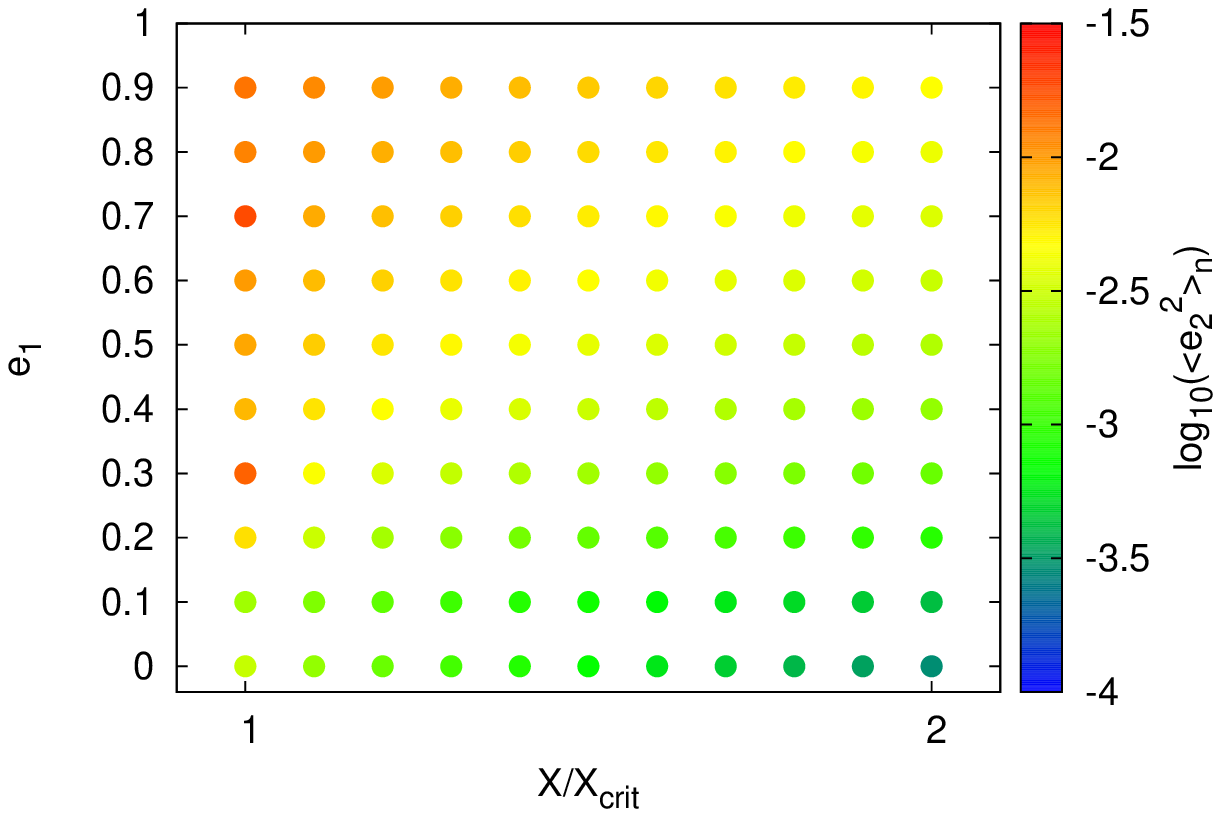}
\includegraphics[width=80mm]{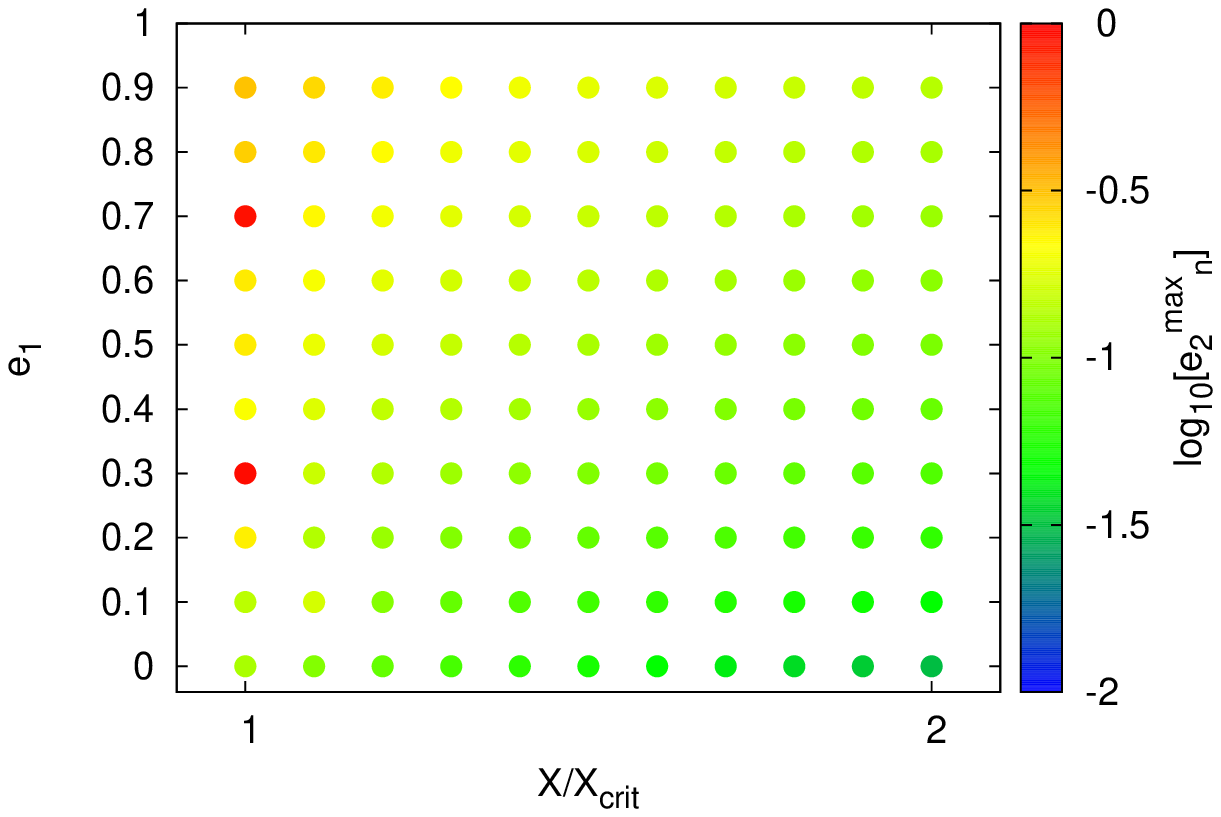}

\caption[]{A zoom in of the plots of Figure \ref{fig:test2} for $1 \leq X/X_{crit}\leq 2$.
\label{fig:test3}}
\end{center}
\end{figure}

\begin{figure}
\begin{center}
\includegraphics[width=80mm]{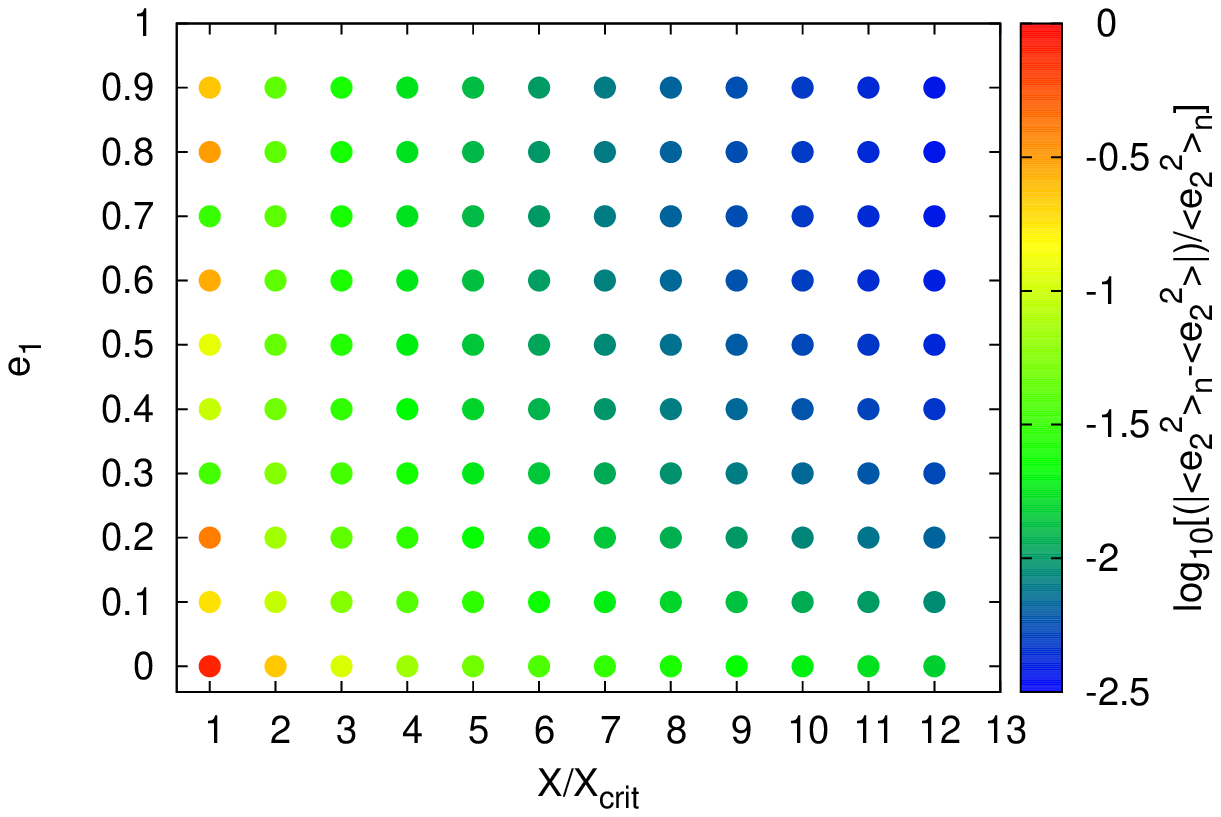}
\includegraphics[width=80mm]{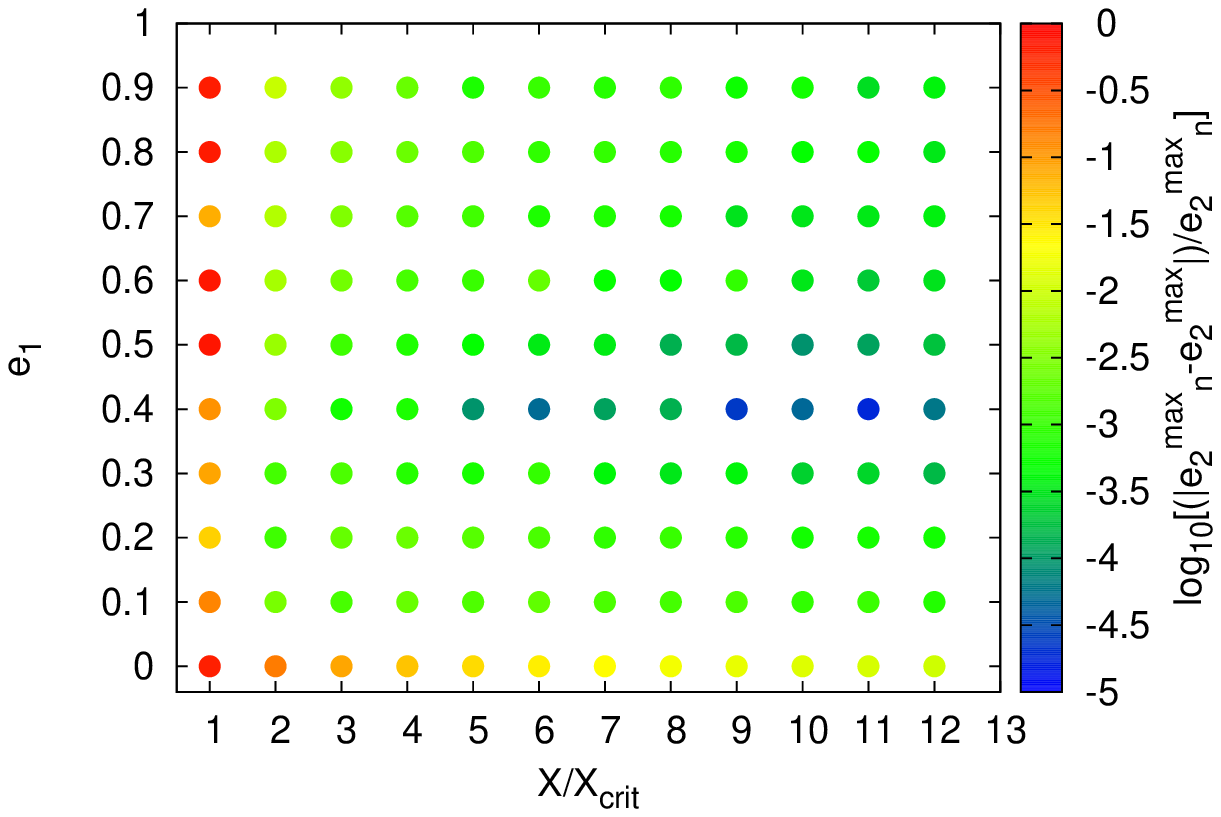}
\includegraphics[width=80mm]{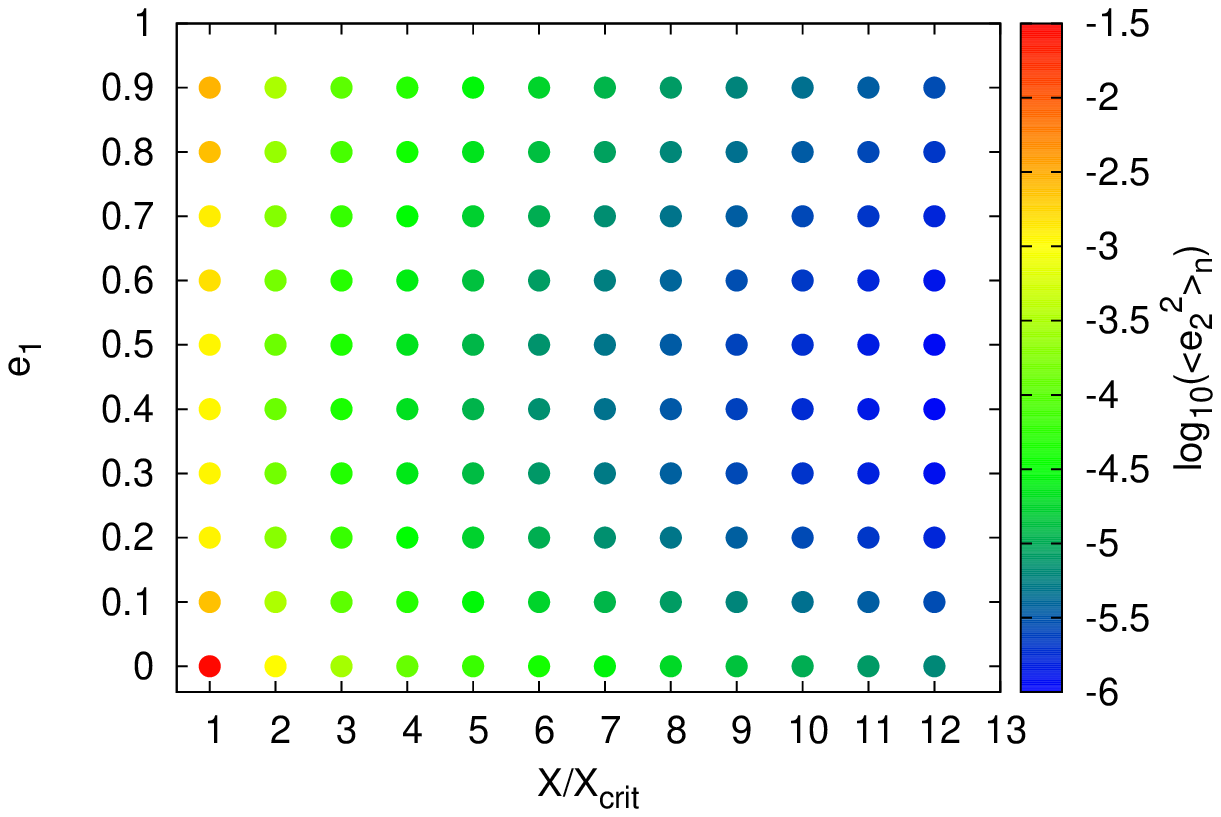}
\includegraphics[width=80mm]{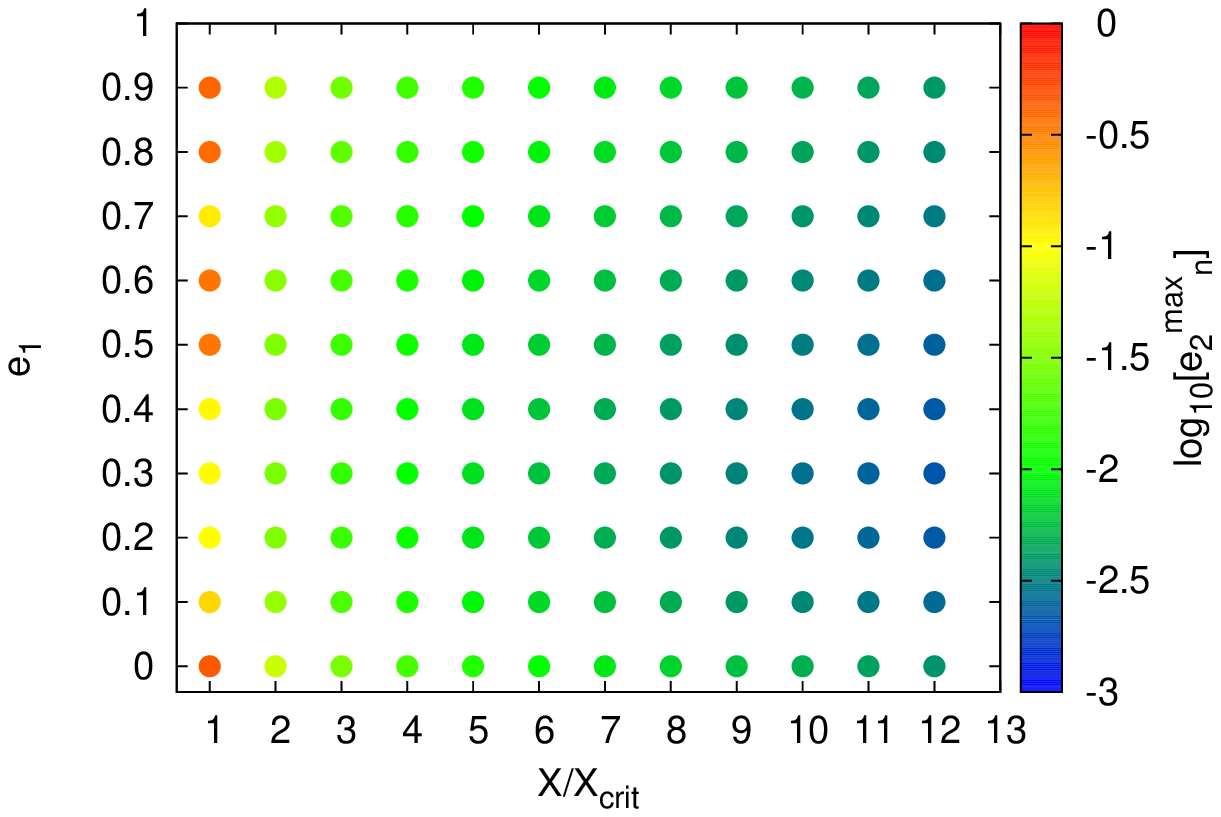}

\caption[]{Top row: logarithmic error (in color) for the averaged square planetary eccentricity (left) and the maximum planetary eccentricity (right).  Bottom row: the logarithms of the numerical values (in colour) of the averaged squared planetary eccentricity (left) and the maximum planetary eccentricity (right).
The mass parameters of the system are $m_1/(m_0+m_1)=0.1$ and $m_2/(m_0+m_1)=10^{-6}$, $X$ is the initial period ratio and $X_{crit}$ is the critical period ratio based on \citet{1999AJ....117..621H}.  The integration time is one analytical secular period.
\label{fig:test4}}
\end{center}
\end{figure}

\begin{figure}
\begin{center}
\includegraphics[width=80mm]{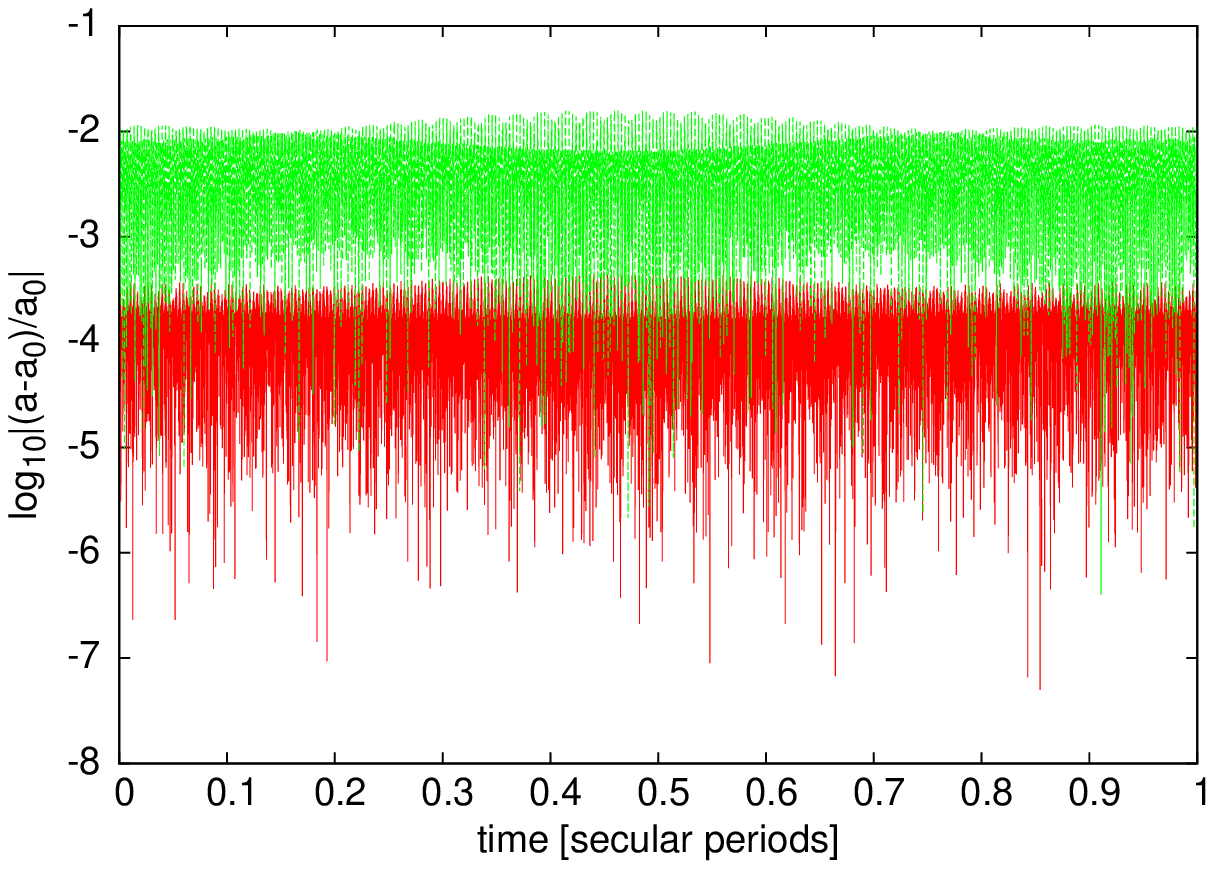}
\includegraphics[width=80mm]{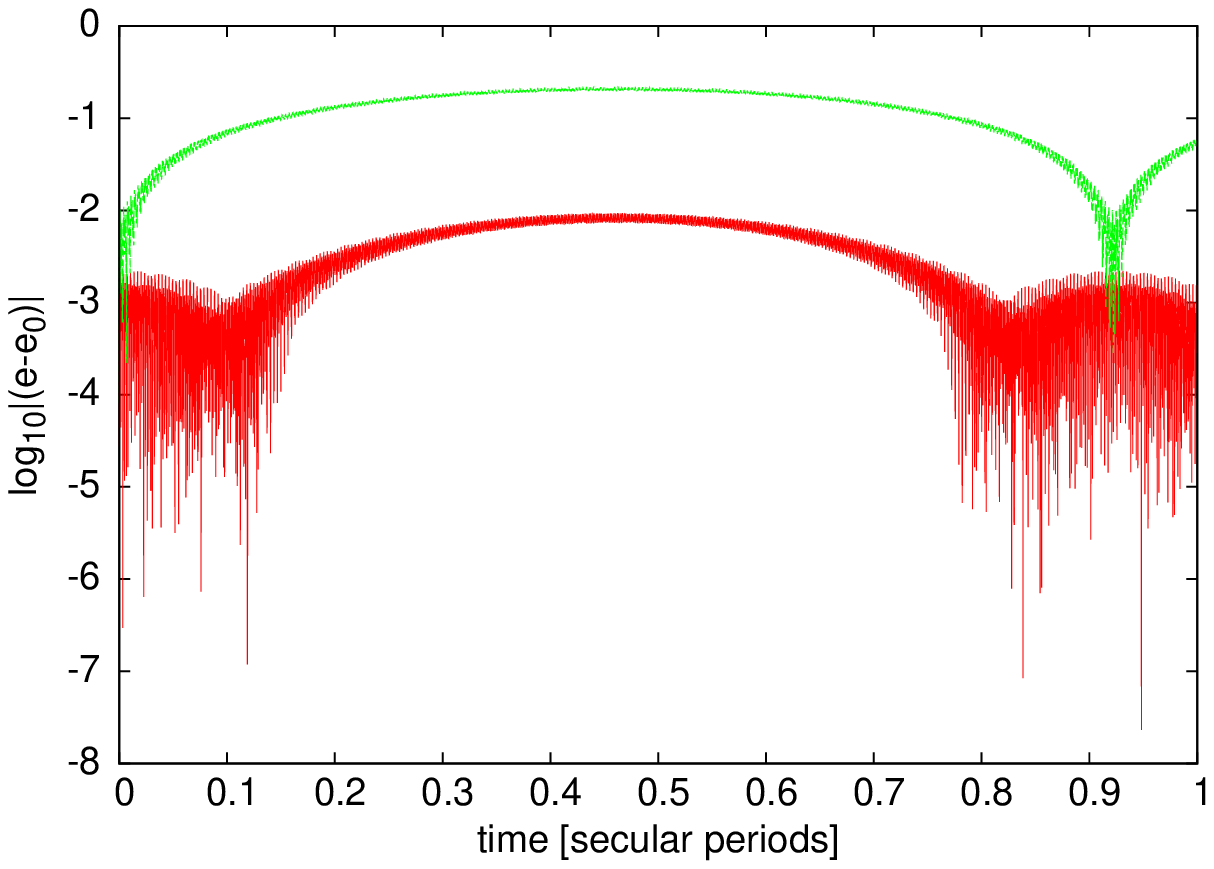}
\caption[]{Semi-major axis logarithmic relative change (left graph) and logarithmic absolute change in the eccentricity (right graph). 
The red colour represents the stellar orbit, while the green colour represents the planetary orbit. 
The mass parameters of the system are $m_1/(m_0+m_1)=0.1$ and $m_2/(m_0+m_1)=10^{-2}$, the initial stellar eccentricity is 0.5, 
the initial stellar eccentric anomaly is $0^{\circ}$, the initial planetary mean anomaly is $90^{\circ}$, 
the initial period ratio is $1.5 X_{crit}$ and the integration time is one analytical secular period.  
Variations in the 
planetary eccentricity remain dominant even for systems close to the critical period ratio for dynamical instability.
\label{fig:test5}}
\end{center}
\end{figure}

Generally, our analytical estimates were in good agreement with the results obtained from the numerical simulations.  As expected, the 
greatest errors were for systems where the planet was started at the critical semi-major axis.  As we move away from the critical
semi-major axis the error reduces rapidly.  Figures \ref{fig:test1}-\ref{fig:test4} show the most relevant numerical findings.  Finally,
Figure \ref{fig:test5} is an example that demonstrates the validity of our assumption that the semi-major axes and stellar eccentricity 
remain constant.

\section{POST-NEWTONIAN CORRECTION}
\label{sec:PN}
\begin{figure}
\begin{center}
\includegraphics[width=90mm]{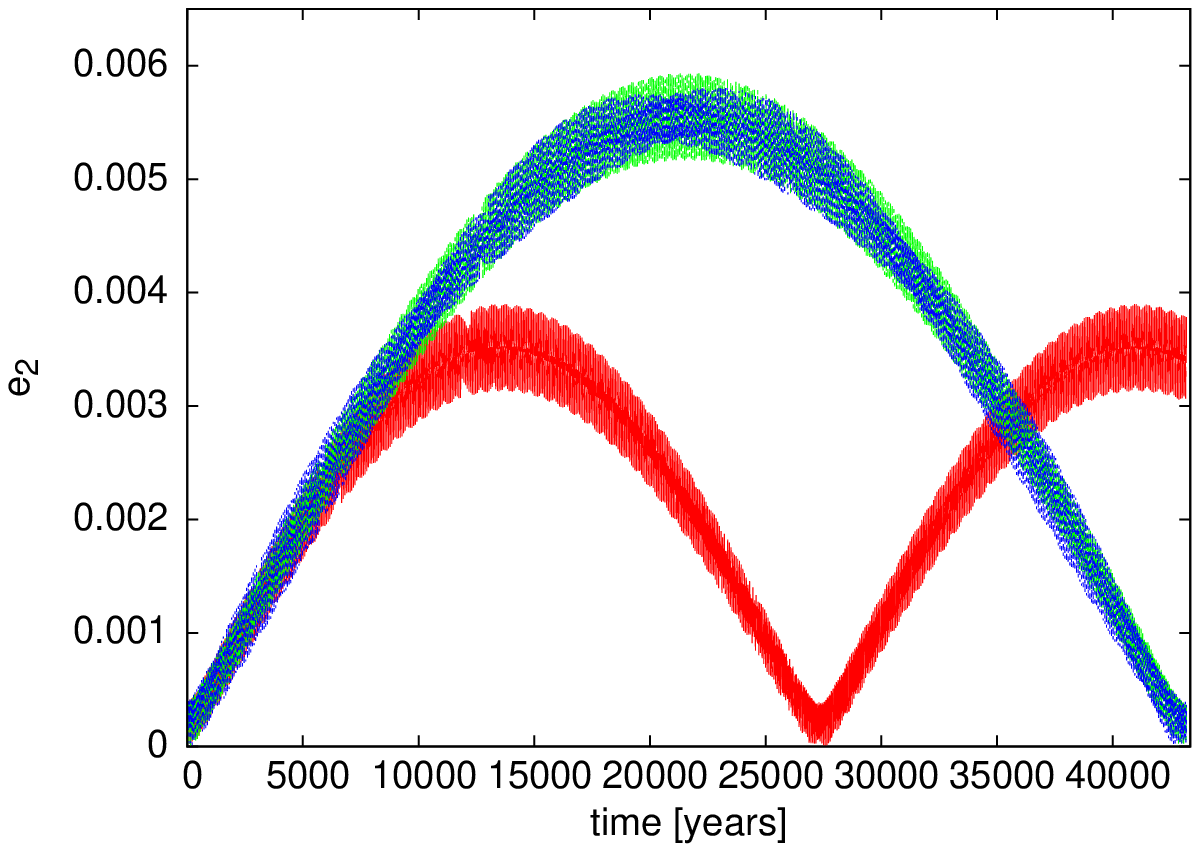}
\caption[]{Eccentricity against time for for a system with $m_0=5 M_{\odot}$, $m_1=4 M_{\odot}$, $m_2=1 M_J$, $a_1=0.2$ au, $a_2=6.84014558$ au, $e_1=0.5$, $\varpi_1=0^{\circ}$, $E_{10}=0^{\circ}$ and $l_{20}=90^{\circ}$. The red curve comes from the integration of the non-relativistic full equations of motion, the green curve comes from the integration of the relativistic full equations of motion and the blue curve is our analytic estimate.}
\label{fig:PN}
\end{center}
\end{figure}

For certain stellar binaries the gravitational field between the two stars can be strong enough to make it necessary to include
General Relativity in order to describe the motion of the system correctly. In such cases, our theory can easily be modified to account for the extra perturbation.  
One simply adds the following first order post-Newtonian correction for the stellar orbit to the equation (\ref{ham}) \citep[e.g.][]{Naoz-et-al-2013b}:

\begin{equation}
H_{1PN}=-\frac{(m^3_0+m^3_1)p^4}{8c^2m^3_0m^3_1}-\frac{\mathcal{G}(3m^2_0+7m_0m_1+3m^2_1)p^2}{2c^2m_0m_1r}-
\frac{\mathcal{G}(\textit{\textbf{p}}\cdot\textit{\textbf{r}})^2}{2c^2r^3}+\frac{\mathcal{G}^2(m_0+m_1)^2m}{2c^2r^2},
\end{equation}
where ${\textit{\textbf{p}}}$ and p are the linear momentum and its magnitude respectively and ${c}$ is the speed of light in vacuum.
Averaging the above Hamiltonian over the fast motion, we obtain \citep{Naoz-et-al-2013b}:
\begin{equation}
<H_{1PN}>=\frac{\mathcal{G}^4m^5_0m^5_1(15m^2_0+29m_0m_1+15m^2_1)}{8(m_0+m_1)^3c^2L^4_{1l}}-\frac{3\mathcal{G}^4m^5_0m^5_1}{(m_0+m_1)c^2L^3_{1l}G_{1l}}
\end{equation}
which eventually results in having the following extra term:
\begin{equation}
\frac{dg_{1PN}}{dt}=\frac{3\mathcal{G}^{\frac{3}{2}}(m_0+m_1)^{\frac{3}{2}}}{c^2a^{\frac{5}{2}}_{1l}(1-e^2_{1l})}.
\end{equation}
The above term can simply be added to equation (\ref{post}). Then we have
\begin{equation}
\frac{dg_{1}}{dt}\approx\frac{dg_{1l}}{dt}+\frac{dg_{1PN}}{dt}
\end{equation}
and consequently
\begin{equation}
g_{1}(t)\approx K_{3PN}\;t \label{eq:finalg1}
\end{equation}
where 
\begin{equation}
K_{3PN}=\frac{3}{4}\frac{\mathcal{G}^{\frac{1}{2}}m_2a^{\frac{3}{2}}_{1l}(1-e^2_{1l})^{\frac{1}{2}}}{(m_0+m_1)^{\frac{1}{2}}a^3_{2l}}+
\frac{3\mathcal{G}^{\frac{3}{2}}(m_0+m_1)^{\frac{3}{2}}}{c^2a^{\frac{5}{2}}_{1l}(1-e^2_{1l})}.
\end{equation}
An example of the effect of the post-newtonian (PN) correction is given in Figure \ref{fig:PN}. 
Here, we have integrated numerically the Einstein-Infeld-Hoffmann equations using a Gauss-Radau scheme \citep{2010LNP...790..431E} 
for a system with $m_0=5 M_{\odot}$, $m_1=4 M_{\odot}$, $m_2=1 M_J$, $a_1=0.2$ au, $e_1=0.5$, $\varpi=0^{\circ}$, $E_{10}=0^{\circ}$ and $l_{20}=90^{\circ}$.  
The initial period ratio for the above system was X=200, yielding a planetary semi-major axis $a_2=6.84014558$ au.
The graph demonstrates that the inclusion of GR in modelling the motion of the stellar binary 
can have a significant effect on the evolution of the planetary eccentricity.  

The post-Newtonian correction can be applied to our analytic estimates by simply exchanging $K_3$ with $K_{3PN}$, for instance, in equations (\ref{eq:f2}) and (\ref{eq:e2max}) - (\ref{eq:e2av}).

\section{ANALYTIC DESCRIPTION OF THE SYSTEM'S EVOLUTION}
\label{sec:ana}
In this section we demonstrate how equations (\ref{eq:e1l}), (\ref{eq:f2}) and (\ref{eq:finalg1}) can be used to describe the time evolution of the entire system, i.e. the binary
and the planet. 
For the inner orbit we have
\begin{eqnarray}
\varpi_1&\approx&g_{1}=K_{3PN} t,\\
l_1&=& n_1 t + l_{10},\\
r \cos f_1&=&a_1 (\cos{E_1} - e_1),\\
r \sin f_1&=&a_1 (1-e_1^2)^{1/2} \sin{E_1},
\end{eqnarray}
where $f_1$ denotes the corresponding true anomaly. 
Here we have assumed that $\dot{a}_1=\dot{e}_1=0$. For the mean motion follows that $n_1=\mathcal{G}^{1/2}(m_0+m_1)^{1/2}a_1^{-3/2}$.
Defining $\textit{\textbf{W}}_i$ as the simple 2D
rotation matrix 
\begin{eqnarray}
\textit{\textbf {W}}_i =
 \begin{pmatrix}
  \cos \varpi_i & -\sin \varpi_i \\
  \sin \varpi_i & \cos \varpi_i 
 \end{pmatrix}, \qquad i=1,2
\end{eqnarray}
we find the solution for the relative vector of the binary stars, $\vec r(t)$ is 
\begin{equation}
\textit{\textbf{r}}=\textit{\textbf{W}}_1 (r \cos{f_1}, r \sin{f_1})^T.
\end{equation}
In order to find the position vector of the planet, one follows the same procedure, except now the length of the eccentricity vector is no longer constant
in time and we have
\begin{eqnarray}
\varpi_2&=&\arctan (e_{22}/e_{21}),\\
l_2&=& n_2 t + \varpi_2 + l_{20},\\
e_2(t)&=&|\textit{\textbf{e}}_2|(t),\\
R \cos f_2&=&a_2 (\cos{E_2} - e_2),\\
R \sin f_2&=&a_2 (1-e_2^2)^{1/2} \sin{E_2},
\end{eqnarray}
where  $n_2=(\mathcal{G}M)^{1/2}a_2^{-3/2}$. Here, the corresponding values for $\vec{e}_2$ have to be taken from equations (\ref{eq:f2}).  The relative vector form the barycenter of the binary star to the planet is then given by
\begin{equation}
\textit{\textbf{R}}=\textit{\textbf{W}}_2(R \cos{f_2}, R \sin{f_2})^T.
\end{equation}
Again, we have assumed that $\dot{a}_2=0$.
We would like to point out here that the eccentric anomaly $E_i$ for any of the two orbits can be computed by solving Kepler's equation $l_i=E_i-e_i \sin{E_i}$.  Alternatively, a series expansion that relates the mean and the true anomaly could be used, for instance \citep[e.g see][]{1999ssd..book.....M}:
\begin{equation}
f_i=l_i+2e_i\sin{l_i}+\frac{5}{4}e_i^2\sin{2l_i}+O(e_i^3)
\end{equation}
However, one should keep in mind that the above series solution is divergent for $e_i>0.6627434$. 

A transformation to barycentric coordinates can be easily achieved through
\begin{eqnarray}
\textit{\textbf{r}}_{0b}&=&-\frac{m_1}{m_0+m_1}\textit{\textbf{r}}-\frac{m_2}{M}\textit{\textbf{R}}\\
\textit{\textbf{r}}_{1b}&=&\frac{m_0}{m_0+m_1}\textit{\textbf{r}}-\frac{m_2}{M}\textit{\textbf{R}}\\
\textit{\textbf{r}}_{2b}&=&\frac{m_1+m_2}{M}\textit{\textbf{R}},
\end{eqnarray}
where $\textit{\textbf{r}}_{ib}$ is the vector of the $m_i$ body with respect to the system's barycentre.

\section{APPLICATIONS TO REAL SYSTEMS}
\label{sec:kepler}
We now apply our analytical theory to real circumbinary planetary systems. For that purpose, 
the Kepler-16, Kepler-34, Kepler-35, Kepler-38, Kepler-64 and Kepler-413 systems were selected, as they are currently believed to harbour only one planet in a circumbinary orbit.  The systems are assumed to be coplanar, $\varpi_1=0^{\circ}$, $E_{10}=0^{\circ}$ and $l_{20}=90^{\circ}$, while
the rest of the system parameters were taken from the corresponding discovery papers \citep{2011Sci...333.1602D, Welsh-et-al-2012, 2012ApJ...758...87O, 2013ApJ...768..127S, 2014ApJ...784...14K}.  The systems were integrated over one analytical secular period and no other effects than Newtonian gravity were considered, as they were not expected to make a significant contribution to the systems under investigation \citep[e.g.][]{2015MNRAS.446.1283C}.  Table 1 gives the mass parameters and orbital elements of each system.  

\begin{figure}
\begin{center}
\includegraphics[width=80mm]{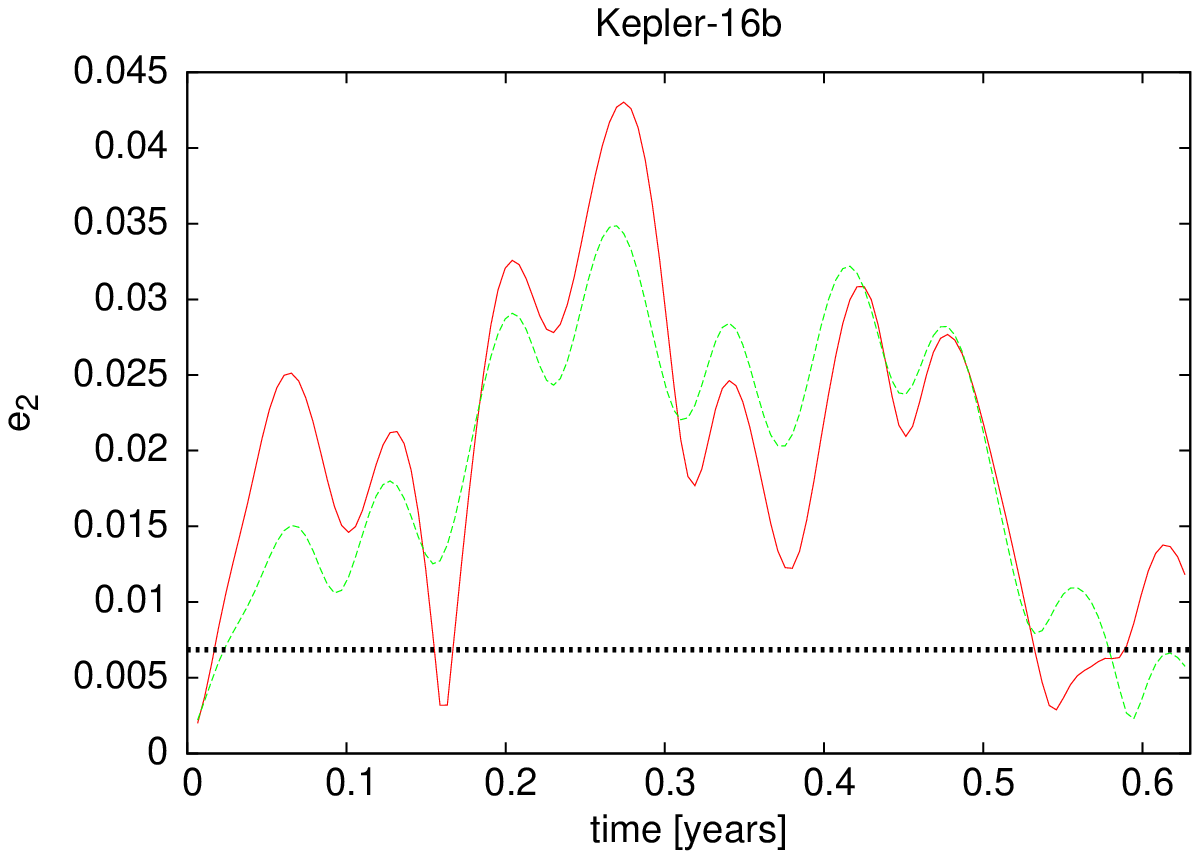}
\includegraphics[width=80mm]{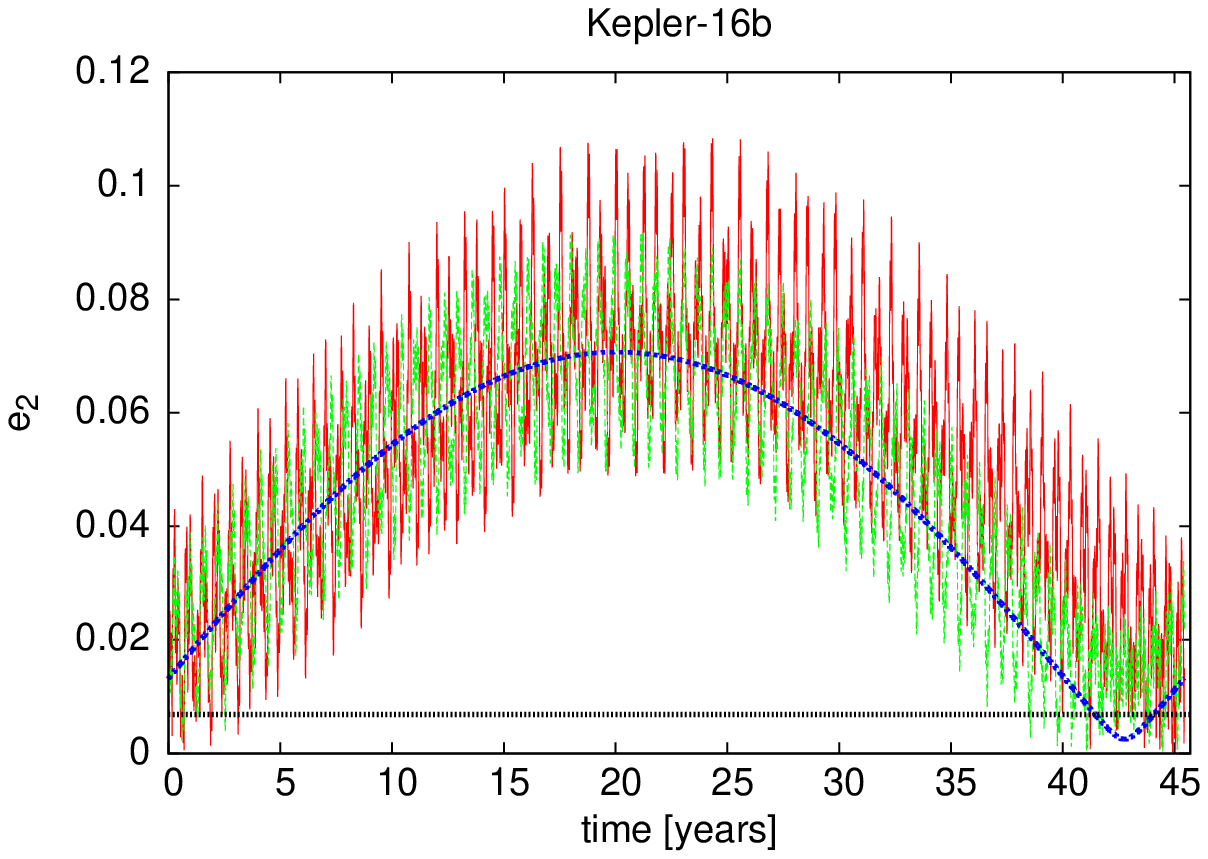}
\includegraphics[width=80mm]{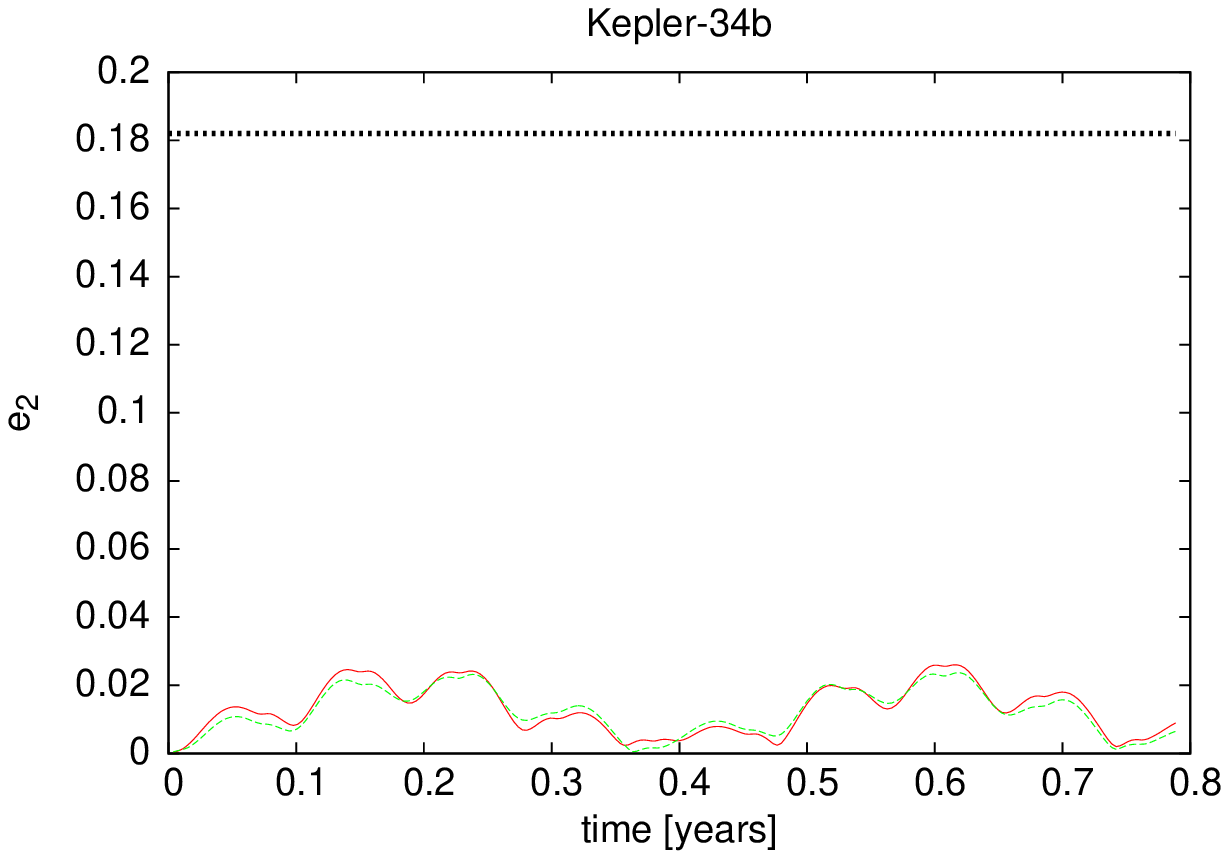}
\includegraphics[width=80mm]{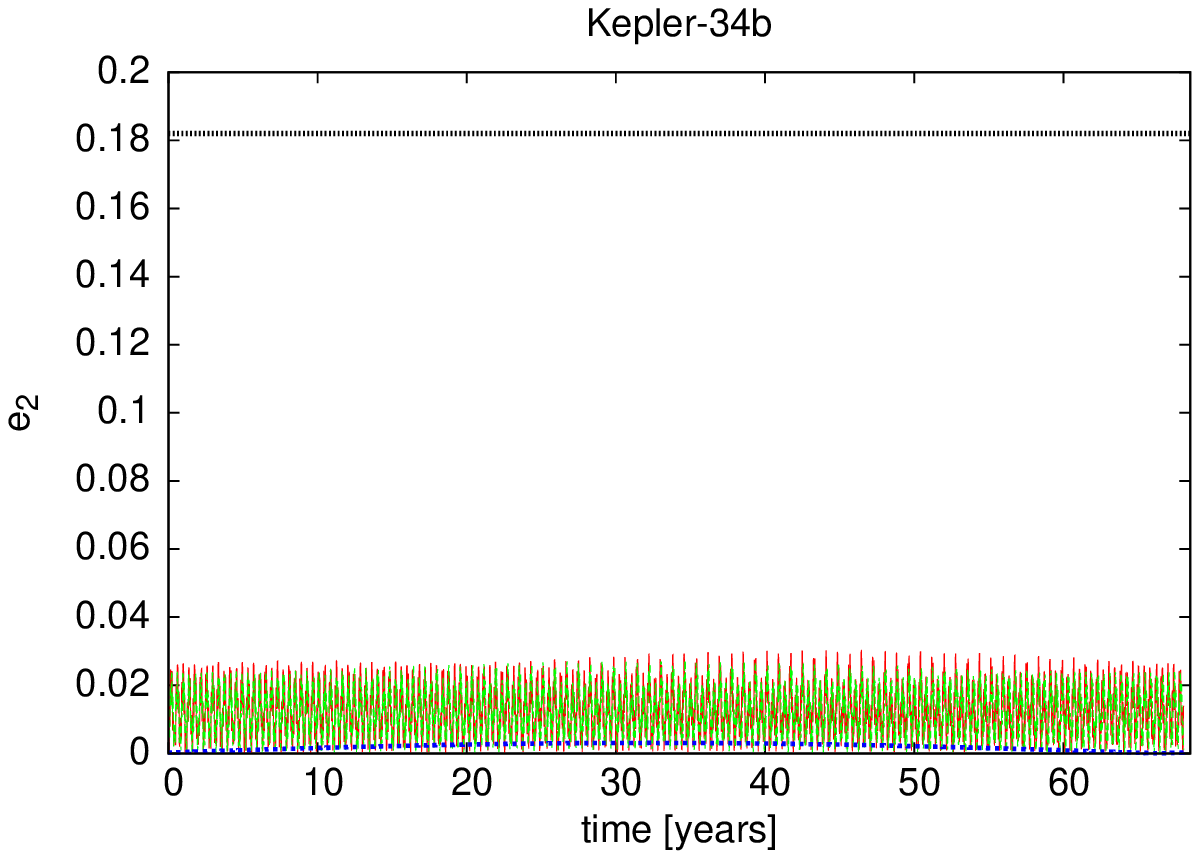}
\includegraphics[width=80mm]{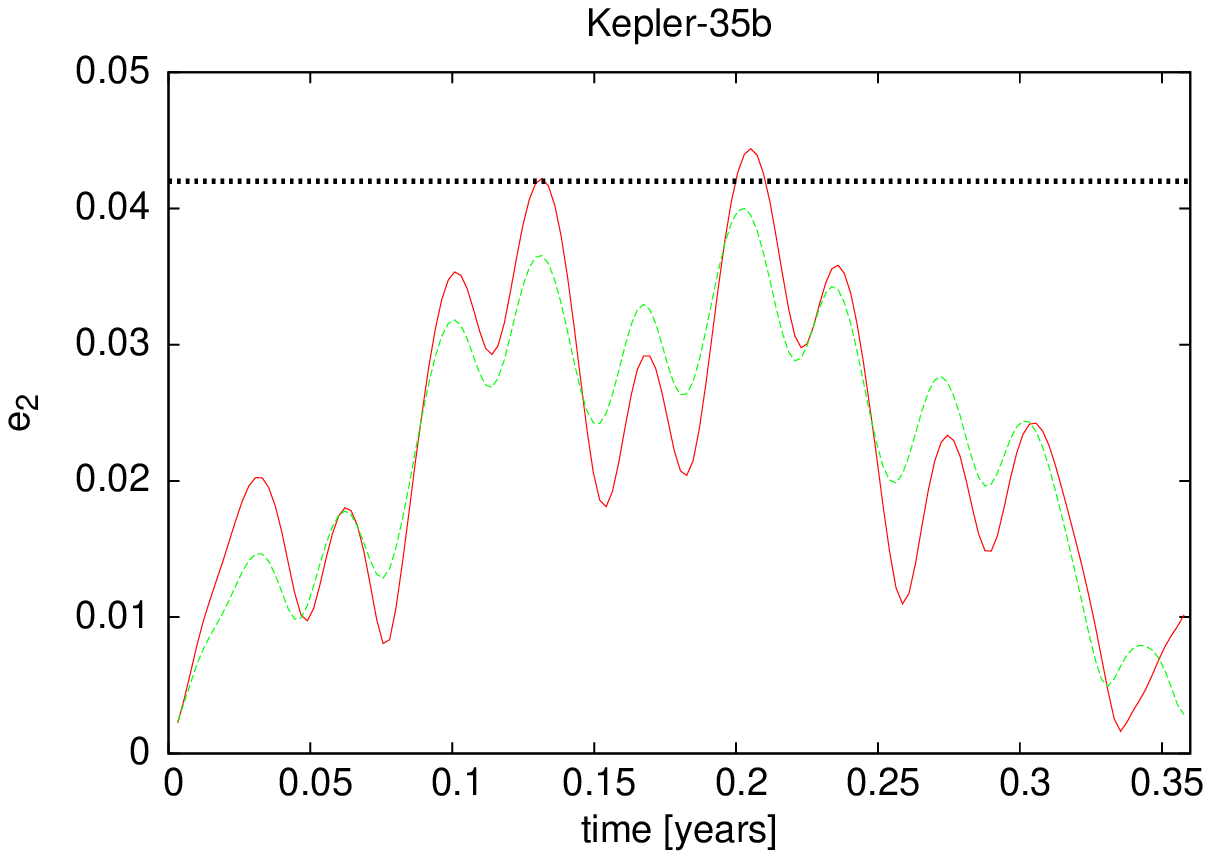}
\includegraphics[width=80mm]{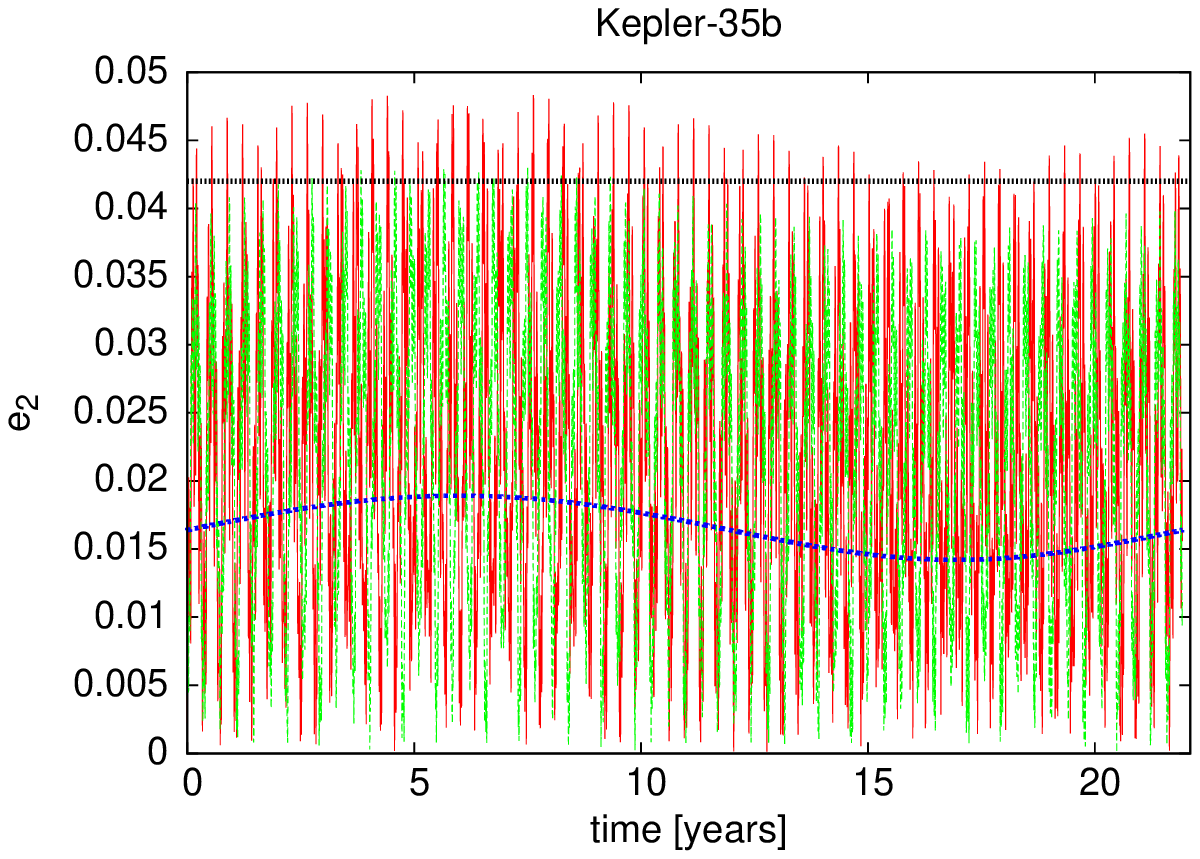}
\caption[]{Eccentricity against time for Kepler-16b, Kepler-34b and Kepler-35b.  The red curve comes from the numerical integration of the full equations of motion, the green curve is our analytical estimates, the blue curve is the analytical secular solution, while the black line denotes the current value of the planetary eccentricity. The integration time is one planetary period for the left column and one analytical secular period for the right column.
\label{fig:kepler1}}
\end{center}
\end{figure}

\begin{figure}
\begin{center}
\includegraphics[width=80mm]{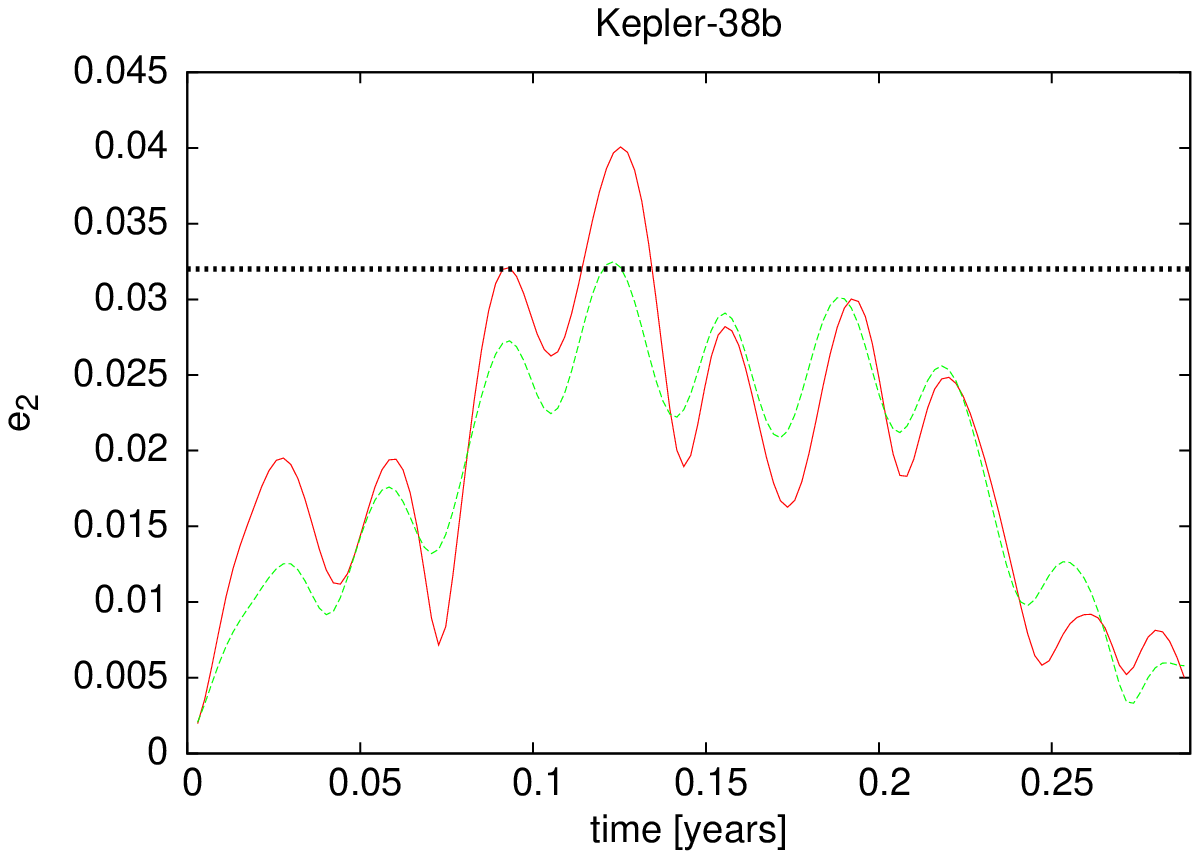}
\includegraphics[width=80mm]{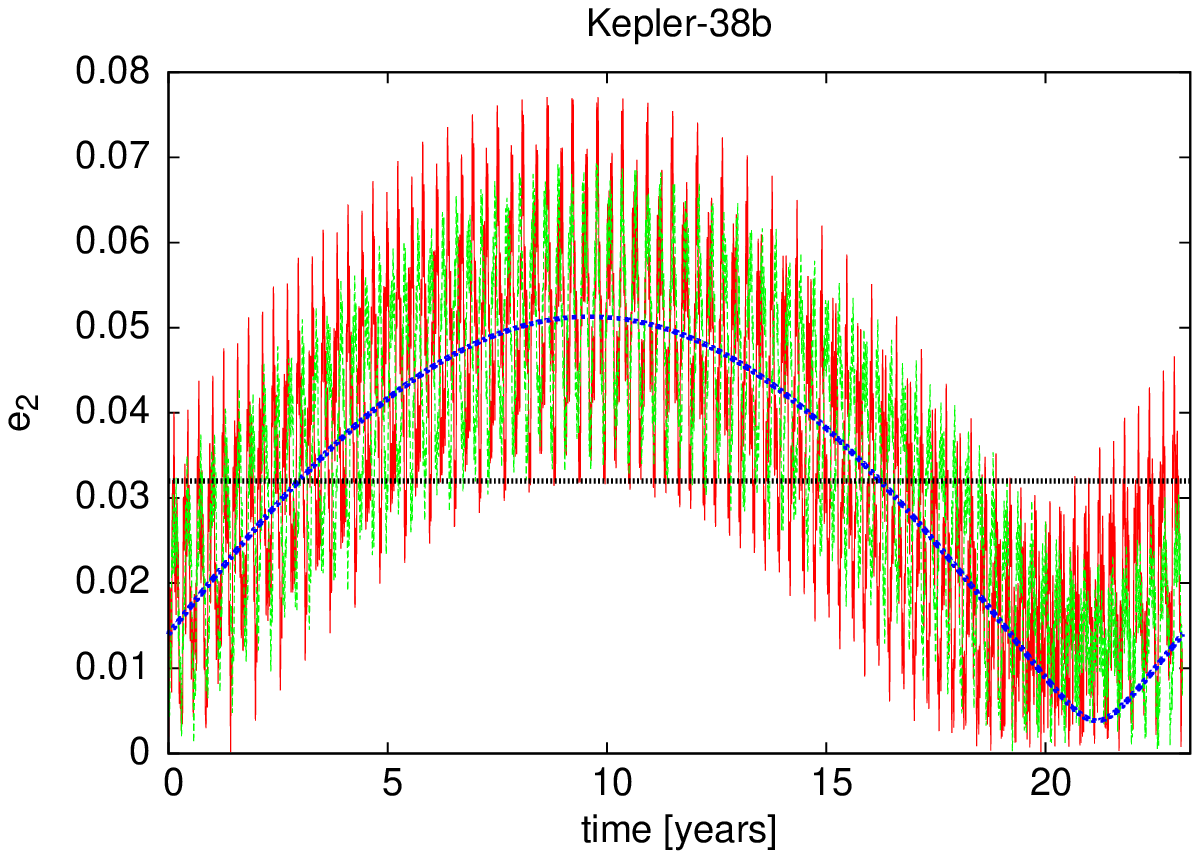}
\includegraphics[width=80mm]{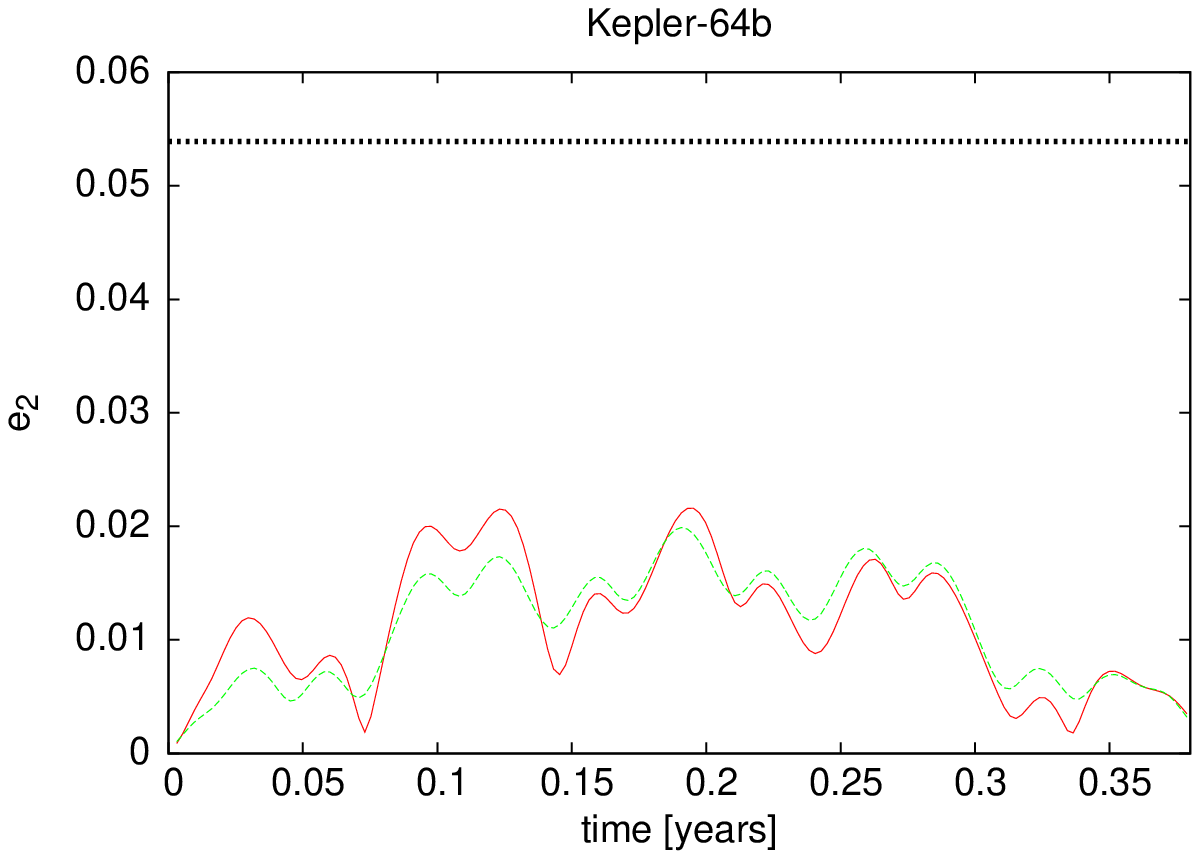}
\includegraphics[width=80mm]{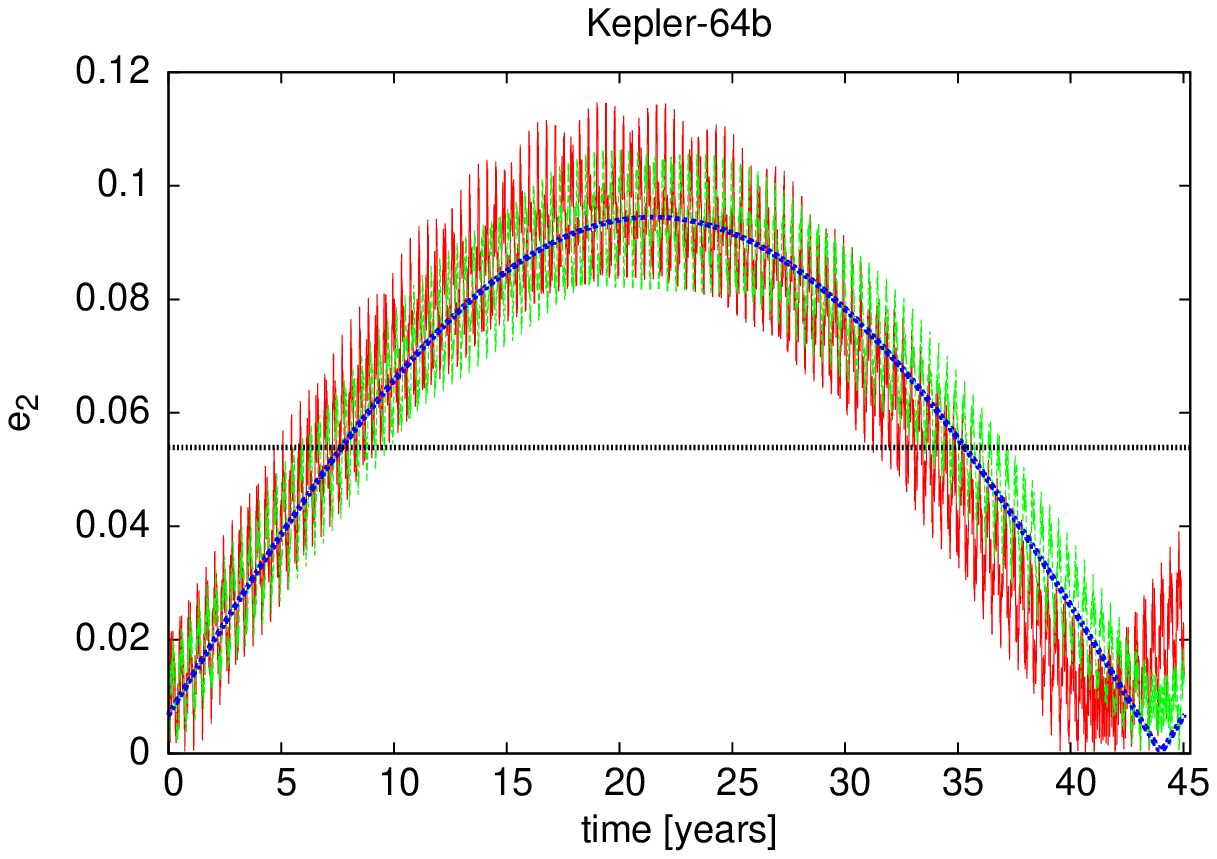}
\includegraphics[width=80mm]{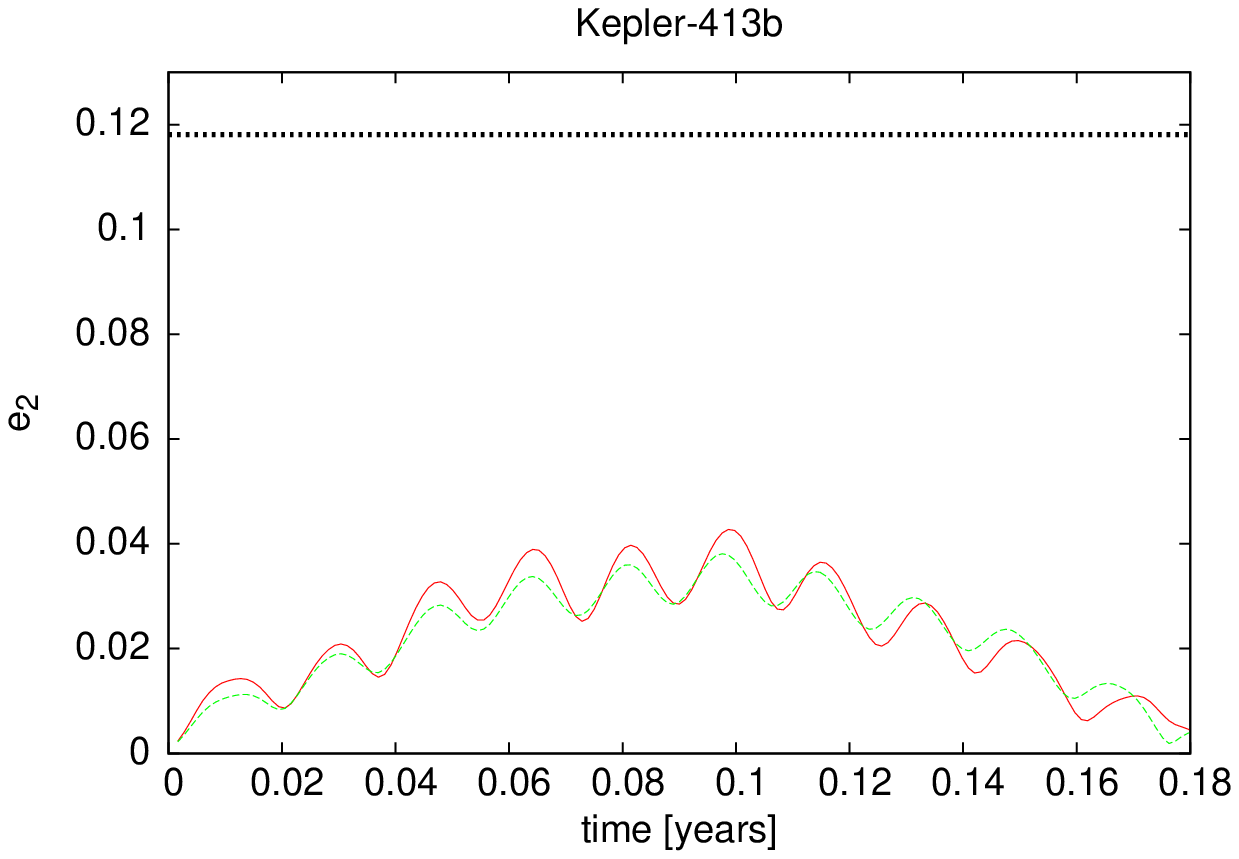}
\includegraphics[width=80mm]{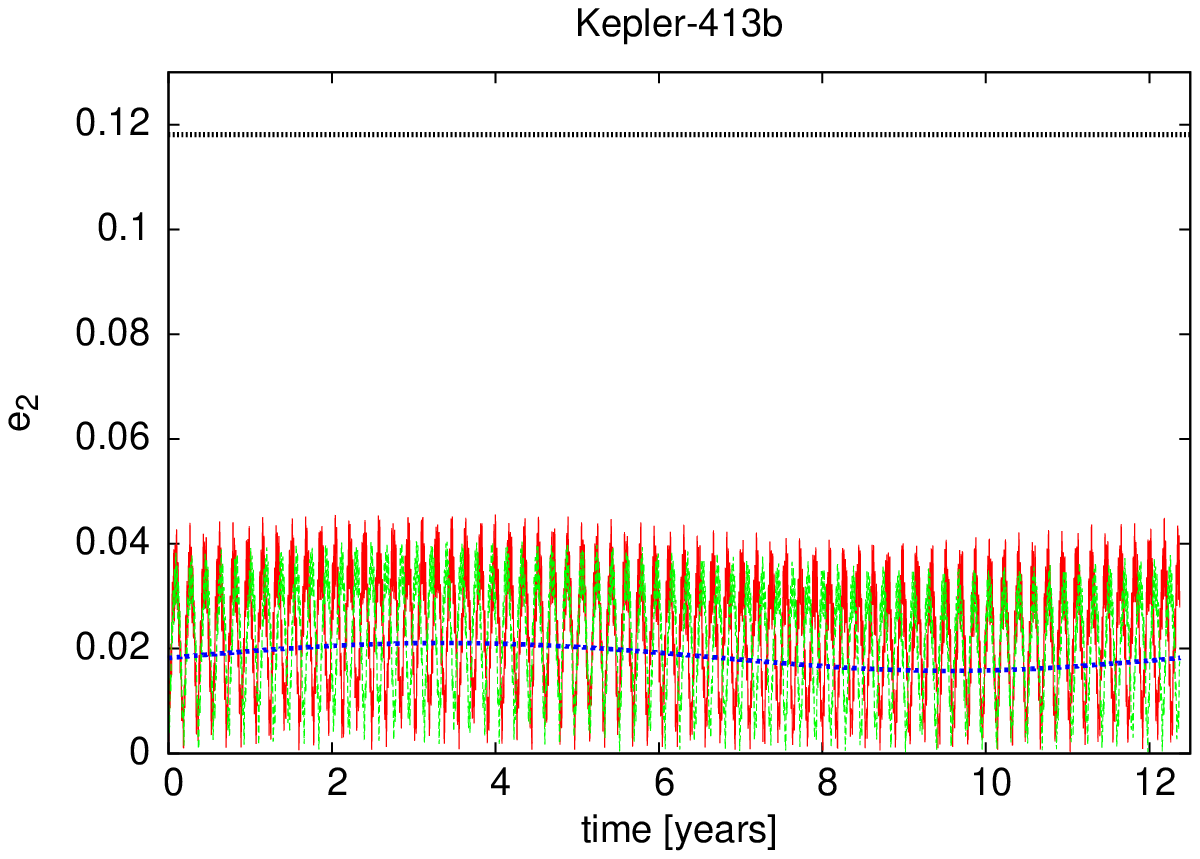}
\caption[]{Same as Figure \ref{fig:kepler1}, but for Kepler-38b, Kepler-64b and Kepler-413b.  \label{fig:kepler2}}
\end{center}
\end{figure}

Figures \ref{fig:kepler1} and \ref{fig:kepler2} show the results for the six Kepler systems.  Generally, the numerical results are in good agreement with the analytical estimates. Furthermore, one can see that for most planets the current state of eccentricities, indicated by a black horizontal line, 
is compatible with formation scenarios that predict initial orbits with low eccentricities after the gaseous phase.
As in-situ planetesimal accretion as well as gravitational collapse have practically been ruled out for most of the circumbinary planets discovered during the Kepler mission
\citep[e.g.][]{2013MNRAS.429..895P, 2014ApJ...782L..11L}, 
a fast disc driven migration with little time 
spent near resonances seems to be the most likely formation scenario for Kepler-16, Kepler-35, Kepler-38 and Kepler-64 \citep[e.g.][]{2014A&A...564A..72K}.

Exceptions are Kepler-34 and Kepler-413, both with a higher planetary eccentricity of $e_2=0.182$ and  $e_2=0.1181$, respectively. 
Looking at the relevant plots, 
it is clear that starting the planet on a circular orbit can not produce a planetary orbit with 
eccentricities higher than 0.03 for Kepler-34b and 0.04 for Kepler-413b.  
Moreover, the main eccentricity contribution for both systems comes from short period activity. This is to be expected, as the stellar
masses of Kepler-34 have only around $2.5\%$ difference, and the stellar eccentricity of 
the Kepler-413 is just 0.0365. As a result, the forced secular eccentricity, which is proportional to the difference between the masses of the stellar components and to the stellar eccentricity is very small.  Therefore, either those two planets were formed 
on a non-circular orbit or if they were initially circular, some dynamical event may have taken place and pumped up their eccentricity.
For instance, an as of yet undetected companion as well as an encounter with another star may have injected eccentricity into the planet's orbit. 
Such an interaction would also explain the slight misalignment of the orbital planes in Kepler-413.  
Another possible explanation for the elevated Kepler-34b is 
resonance trapping. If the planet's migration has not been fast enough, the planet may be trapped in a resonance which 
can cause a significant increase in its orbital eccentricity \citep{2014A&A...564A..72K}. In the case of Kepler-34b the 10:1 mean motion resonace with the stellar binary 
may have affected the evolution of the planetary eccentricity to some extent \citep{2015MNRAS.446.1283C}.

\begin{table}
\begin{scriptsize}
\caption[]{Masses and orbital elements for Kepler-16, Kepler-34, Kepler-35, Kepler-38, Kepler-64 and Kepler-413.} 
\vspace{0.1 cm}
\begin{center}	
{\footnotesize \begin{tabular}{c c c c c c c}\hline
System & $m_0 (M_{\odot})$ & $m_1 (M_{\odot})$ & $m_2 (M_{J})$ & $a_1$ (au) & $a_2$ (au) & $e_1$ \\
\hline
Kepler-16 & $0.6897^{+0.0035}_{-0.0034}$ & $0.20255^{+0.00066}_{-0.00065}$ & $0.333^{+0.016}_{-0.016}$ & $0.22431^{+0.00035}_{-0.00034}$ & $0.7048^{+0.0011}_{-0.0011}$  & 0.$15944^{+0.00061}_{-0.00062}$ \\  
Kepler-34 & $1.0479^{+0.0033}_{-0.0030}$ & $1.0208^{+0.0022}_{-0.0022}$& $0.220^{+0.011}_{-0.010}$ & $0.22882^{+0.00019}_{-0.00018}$ & $1.0896^{+0.0009}_{-0.0009}$  & 0.$52087^{+0.00052}_{-0.00055}$ \\ 
Kepler-35 & $0.8876^{+0.0051}_{-0.0053}$ & $0.8094^{+0.0041}_{-0.0044}$ & $0.127^{+0.020}_{-0.021}$ & $0.17617^{+0.00028}_{-0.00029}$ & $0.60345^{+0.00100}_{-0.00102}$ & $0.1421^{+0.0014}_{-0.0014}$ \\  
Kepler-38 & 0.949 & 0.249 & $<$0.384 ($95\%$ conf.) & 0.1469 & 0.4644 & 0.1032 \\ 
Kepler-64 & $1.384^{+0.079}_{-0.079}$ & $0.386^{+0.018}_{-0.018}$ & $<$0.532 ($99.7\%$ conf.) & $0.1744^{+0.0031}_{-0.0031}$ & $0.634^{+0.011}_{-0.011}$ & 0.$2117^{+0.0051}_{-0.0051}$ \\  
Kepler-413 & 0.820 & 0.5423 & 0.21 & 0.10148 & 0.3553 & 0.0365 \\ 
\end{tabular}}
\end{center}
\end{scriptsize}
\end{table}

\begin{figure}
\begin{center}
\label{fig9}
\includegraphics[width=80mm]{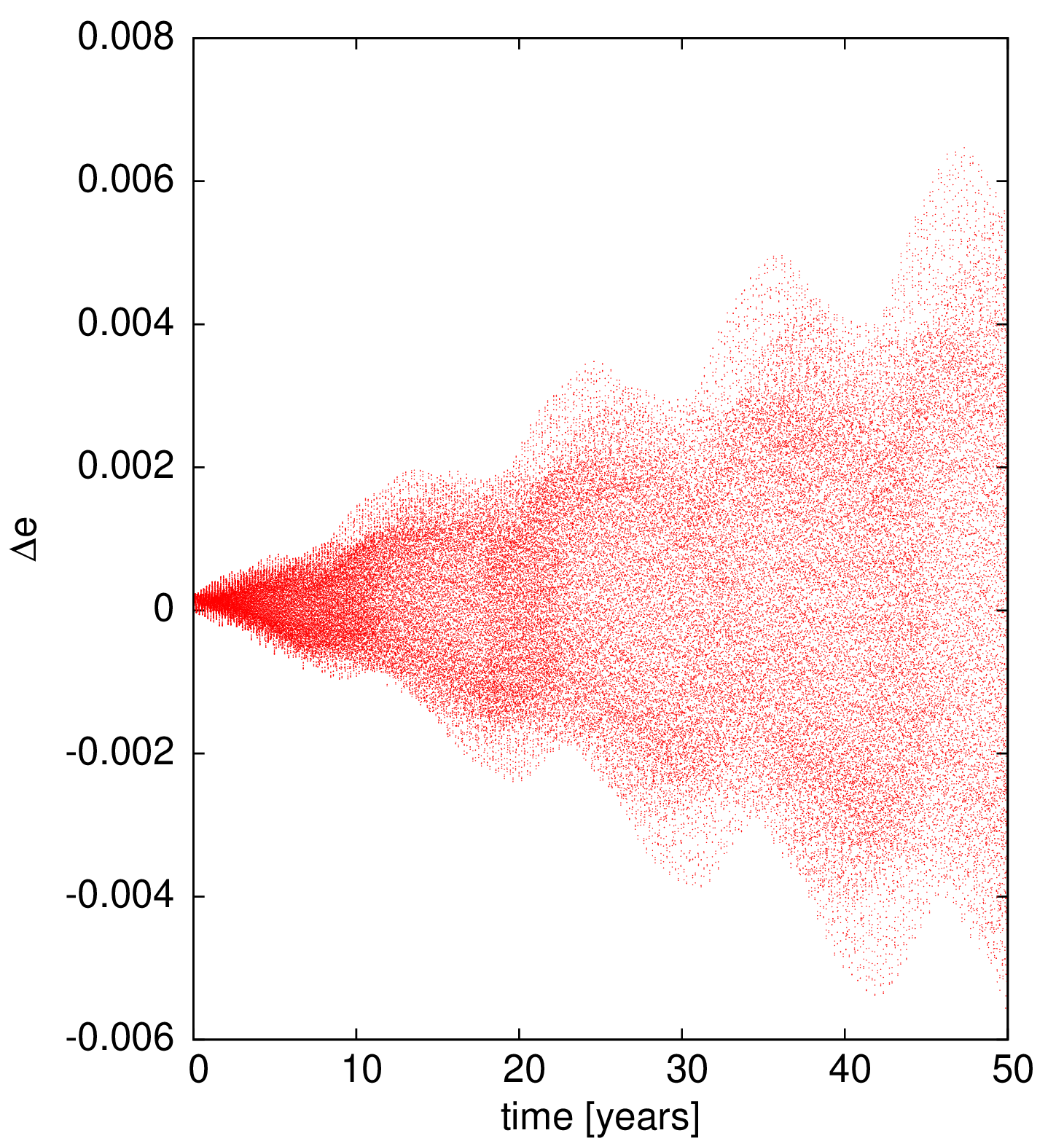}
\includegraphics[width=80mm]{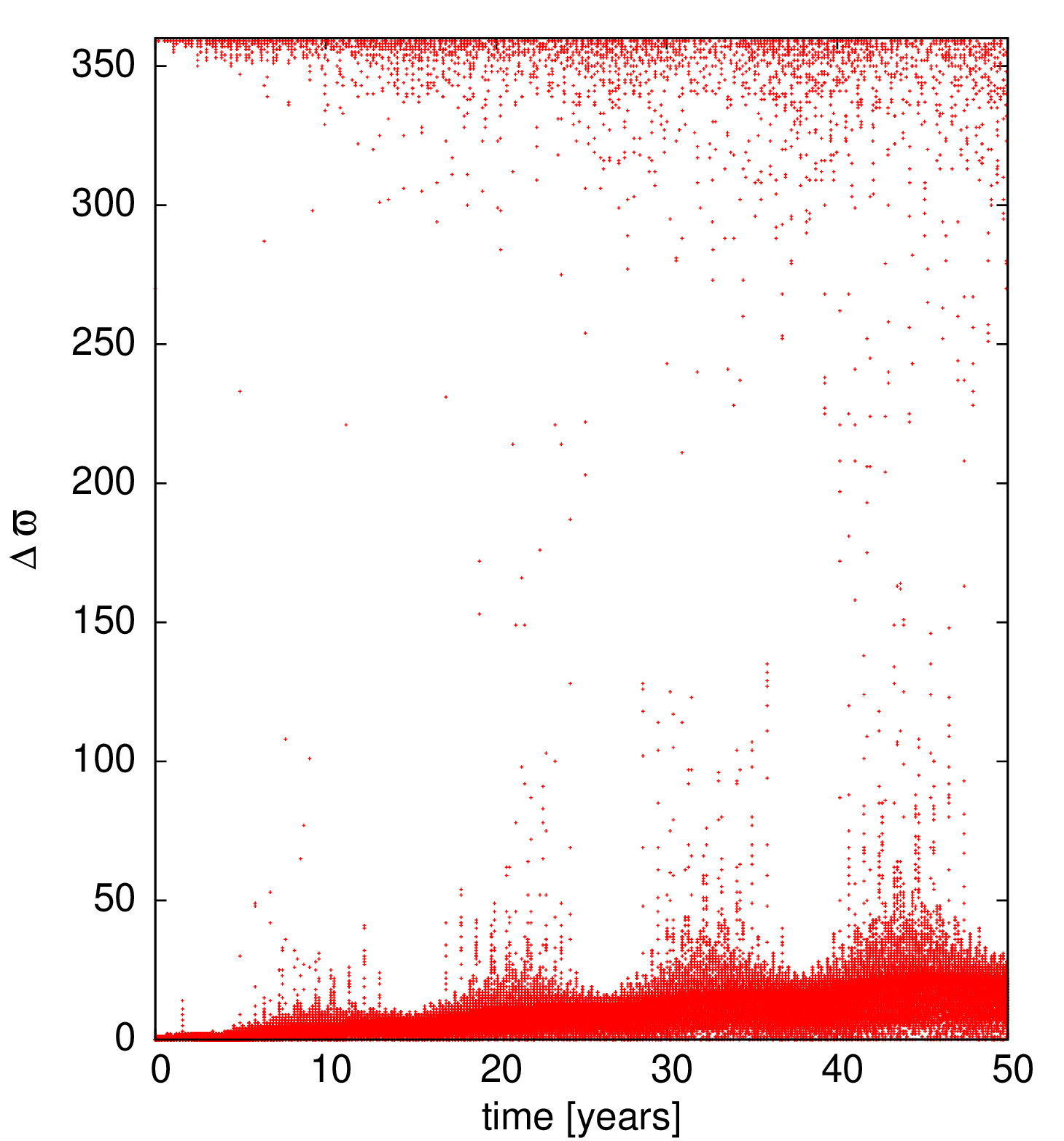}
\caption[]{Difference in the eccentricity (left) and in the longitude of pericentre (right) between a 2D and a 3D model of Kepler-413.
See Figure \ref{fig:kepler2} for a comparison with the current and predicted eccentricity estimates. 
The secular period for a coplanar planetary orbit is about 12 years.}
\end{center}
\end{figure}

\section{DISCUSSION}
We would like to point out here that we assumed our six Kepler systems to be coplanar. In reality, they all have some mutual orbital inclination, with Kepler-413 having the
largest one among those systems, i.e. $I=4.073^{\circ}$. In order to check whether our anayltic description of the system's evolution remains valid for such inclinations,
we ran two and three dimensional simulations for Kepler-413.  We found that difference between the 
two models was insignificant for the amplitude of the eccentricity vector. Therefore, our predictions for the maximum and average eccentricities
remain valid.  However, the longitude of pericentre seemed to be affected considerably. Hence, we do not recommend to use the fully analytic description 
presented in section \ref{sec:ana} to describe a system's long term orbit evolution for mutual inclinations around $I\approx 5^{\circ}$ or higher. Figure 9 presents the relevant information.

\section{SUMMARY}
\label{sec:summary}
In this work, we derived analytical estimates for the motion of coplanar circumbinary planets.  The derivation was done for initially circular orbits and it assumed that
the planetary eccentricity remains low ($e_2<0.2$).  Short term motion was described by solving differential equations derived from the Runge-Lenz vector. Similarly, canonical perturbation theory was used 
to model the secular motion. Wokring with the eccentric instead of the true anomaly to describe the short term motion of the perturber leads 
to a solution that does not depend on a series expansion in terms of the eccentricity of the binary.  This is a great advantage as we do not need to worry about the convergence of the solution and the magnitude of the eccentricity of the perturber.  
Hence, our derivations are valid for all eccentricities of the stellar binary as long as the
planet is far enough from the binary so that a hierarchical triple system can be considered.  

In section \ref{sec:ana} we have presented a complete analytical framework to describe the motion of a stellar binary and its circumbinary planet.
Analytical expressions were derived for the averaged square eccentricity as well as for the maximum value of the planet's eccentricity.  
The estimates were tested numerically for different stellar and planetary masses over a wide range of eccenricity and period ratios.  
The results showed very good agreement between the numerical simulations and the analytical estimates.  
Furthermore, in case it is required, the analytical formulae can easily be adjusted to incorporate some general relativistic effects.  Finally, the analytical estimates
were applied to six Kepler circumbinary systems with good results, as they provided valuable information about the evolution and possibly the formation of those systems.

The analytical estimates obtained in this work can be used, for instance, in analytic planet formation theories, in
modelling transits in circumbinary planetary systems, or determing habitable zones around stellar binaries.

\begin{acknowledgements}
S.E. is grateful for the support by the European Union Seventh Framework Program (FP7/2007-2013) under grant agreement no. 282703, 
as well as the travel funding by the Austrian FWF project S11608-N16 (sub-project of the NFN S116).
\end{acknowledgements} 

\section*{APPENDIX}

The expressions for ${P_1(t)}$ and ${P_2(t)}$ in equations (\ref{eq:s11}) are presented. They contain the short periodic terms
of the outer eccentric vector. For $t=t_0$ we have ${E_1=E_{10}}$ and ${l_2=l_{20}}$.

\begin{eqnarray}
P_1(t) & = &\frac{21}{32}(1-\sqrt{1-e^2_1})\cos{(2E_1+3l_2)}+\frac{3}{32}(1-\sqrt{1-e^2_1})\cos{(2E_1+l_2)}-\frac{3}{32}(1+\nonumber\\
& & +\sqrt{1-e^2_1})\cos{(2E_1-l_2)}-\frac{21}{32}(1+\sqrt{1-e^2_1})\cos{(2E_1-3l_2)}+e_1\bigg[-\frac{21}{96}(1-\nonumber\\
& & - \sqrt{1-e^2_1})\cos{(3E_1+3l_2)}-\frac{3}{96}(1-\sqrt{1-e^2_1})\cos{(3E_1+l_2)}+\frac{3}{32}(13+\nonumber\\
& & +5\sqrt{1-e^2_1})\cos{(E_1-l_2)}+\frac{3}{96}(1+\sqrt{1-e^2_1})\cos{(3E_1-l_2)}-\frac{105}{32}(1-\nonumber\\
& & -\sqrt{1-e^2_1})\cos{(E_1+3l_2)}+\frac{21}{96}(1+\sqrt{1-e^2_1})\cos{(3E_1-3l_2)}-\frac{3}{32}(13-\nonumber\\
& & -5\sqrt{1-e^2_1})\cos{(E_1+l_2)}+\frac{105}{32}(1+\sqrt{1-e^2_1})\cos{(E_1-3l_2)}\bigg]+ e^2_1\bigg[\frac{3}{32}(\frac{7}{2}-\nonumber\\
& &-\sqrt{1-e^2_1})\cos{(2E_1+l_2)}+\frac{21}{32}(\frac{1}{2}-\sqrt{1-e^2_1})\cos{(2E_1+3l_2)}-\frac{3}{32}(\frac{7}{2}+\nonumber\\
& & +\sqrt{1-e^2_1})\cos{(2E_1-l_2)}-\frac{21}{32}(\frac{1}{2}+\sqrt{1-e^2_1})\cos{(2E_1-3l_2)}\bigg]+e^3_1\bigg[-\frac{1}{64}\times\nonumber\\
& & \times\cos{(3E_1+l_2)}-\frac{7}{64}\cos{(3E_1-3l_2)}+\frac{3}{64}\cos{(E_1-l_2)}+\frac{7}{64}\cos{(3E_1+3l_2)}+\nonumber\\
& & +\frac{21}{64}\cos{(E_1+3l_2)}-\frac{3}{64}\cos{(E_1+l_2)}+\frac{1}{64}\cos{(3E_1-l_2)}-\nonumber\\
& & -\frac{21}{64}\cos{(E_1-3l_2)}\bigg]
\end{eqnarray}

\begin{eqnarray}
P_2(t) & = &\frac{21}{32}(1-\sqrt{1-e^2_1})\sin{(2E_1+3l_2)}-\frac{3}{32}(1-\sqrt{1-e^2_1})\sin{(2E_1+l_2)}-\frac{3}{32}(1+\nonumber\\
& & +\sqrt{1-e^2_1})\sin{(2E_1-l_2)}+\frac{21}{32}(1+\sqrt{1-e^2_1})\sin{(2E_1-3l_2)}+
e_1\bigg[-\frac{7}{32}(1-\nonumber\\
& & -\sqrt{1-e^2_1})\sin{(3E_1+3l_2)}+\frac{1}{32}(1-\sqrt{1-e^2_1})\sin{(3E_1+l_2)}-\frac{3}{32}(3-\nonumber\\
& & -5\sqrt{1-e^2_1})\sin{(E_1-l_2)}+\frac{1}{32}(1+\sqrt{1-e^2_1})\sin{(3E_1-l_2)}-\frac{105}{32}(1-\nonumber\\
& & -\sqrt{1-e^2_1})\sin{(E_1+3l_2)}-\frac{7}{32}(1+\sqrt{1-e^2_1})\sin{(3E_1-3l_2)}-\frac{3}{32}(3+\nonumber\\
& & +5\sqrt{1-e^2_1})\sin{(E_1+l_2)}-\frac{105}{32}(1+\sqrt{1-e^2_1})\sin{(E_1-3l_2)}\bigg]+e^2_1\bigg[\frac{3}{32}(\frac{5}{2}+\nonumber\\
& & +\sqrt{1-e^2_1})\sin{(2E_1+l_2)}+\frac{21}{32}(\frac{1}{2}-\sqrt{1-e^2_1})\sin{(2E_1+3l_2)}+\frac{3}{32}(\frac{5}{2}-\nonumber\\
& & -\sqrt{1-e^2_1})\sin{(2E_1-l_2)}+\frac{21}{32}(\frac{1}{2}+\sqrt{1-e^2_1})\sin{(2E_1-3l_2)}\bigg]
+e^3_1\bigg[-\frac{3}{64}\times\nonumber\\
& & \times\sin{(3E_1+l_2)}+\frac{7}{64}\sin{(3E_1-3l_2)}-\frac{9}{64}\sin{(E_1-l_2)}+\frac{7}{64}\sin{(3E_1+3l_2)}+\nonumber\\
& & +\frac{21}{64}\sin{(E_1+3l_2)}-\frac{9}{64}\sin{(E_1+l_2)}-\frac{3}{64}\sin{(3E_1-l_2)}+\nonumber\\
& & +\frac{21}{64}\sin{(E_1-3l_2)}\bigg]
\end{eqnarray}

\section*{NOTATION}
\begin{scriptsize}
\begin{tabular}{ll}

$a_i$ & semimajor axis of the inner ($i=1$) and outer ($i=2$) orbit\\
$c$ & vacuum light speed \\
$C$ & integration constants\\
$\textit{\textbf{e}}_i$ & eccentricity vectors\\
$e_{ij}$ & x ($j=1$) and y ($j=2$) components of the inner and outer eccentricity vector\\
$E_i$ & eccentric anomalies\\
$f_i$ & true anomalies\\
$\mathcal G$ & gravitational constant\\
$\textit{\textbf{h}}$ & specific angular momentum of the outer orbit\\
$H$ & Hamiltonians\\
$I$ & mutual inclination of the binary and planetary orbits\\
$K_i$ & parameters of the long term solution \\
$L_i,G_i$ & Delaunnay actions\\
$l_i,g_i$ & Delaunnay angles (mean anomaly, argument of pericenter)\\
$m_0, m_1$     & masses of the two stars\\
$m_2$ & mass of the planet\\
$M$ & total mass of the system\\
$m,\mathcal M, M_j,\mu_i$ & mass parameters\\ 
$n_i$ & mean motions \\
$\mathcal P_n$ & Legendre polynomials\\
$P_i$ & x and y component terms of short period $\vec e_2$ solution \\
$\textit{\textbf{r}}_{ib}$ & barycentric position vectors of the stars ($i=0,1$) and the planet ($i=2$) \\
$\textit{\textbf{r}}$ & Jacobi vector of the stellar orbit \\
$\textit{\textbf{R}}$ & Jacobi vector of the planetary orbit\\
$\theta$ & angle between $\textit{\textbf{r}}$ and $\textit{\textbf{R}}$\\
$t$ & time \\
$\varpi_i$ & stellar and planetary longitude of pericentre\\
$\textit{\textbf{W}}_i$ & pericentre rotation matrices\\
$X$ & period ratio between outer and inner orbit \\
\end{tabular}
\end{scriptsize}

\bibliographystyle{apj}
\bibliography{ref}

\end{document}